\documentclass[conference]{IEEEtran}
\IEEEoverridecommandlockouts
\usepackage{cite}
\usepackage{amsmath,amssymb,amsfonts}
\usepackage{algorithmic}
\usepackage{graphicx}
\usepackage{textcomp}
\usepackage{xcolor}
\def\BibTeX{{\rm B\kern-.05em{\sc i\kern-.025em b}\kern-.08em
    T\kern-.1667em\lower.7ex\hbox{E}\kern-.125emX}}

\usepackage{arydshln}
\usepackage{minted}
\usepackage{listings}
\usepackage[hidelinks]{hyperref}
\usepackage{cleveref}
\usepackage{enumitem}
\usepackage[dvipsnames]{xcolor}
\usepackage{tikz}
\usetikzlibrary{automata, positioning}
\usepackage{booktabs}
\usepackage{bm}
\usepackage{doi}

\ifCLASSOPTIONcompsoc
    \usepackage[caption=false, font=normalsize, labelfont=sf, textfont=sf]{subfig}
\else
\usepackage[caption=false, font=footnotesize]{subfig}
\fi

\definecolor{identifier}{rgb}{0, 0.1, 0.4}

\lstMakeShortInline[columns=fixed]|
\lstdefinelanguage{spa}{
  keywords={stream, completion, await, relative_stream, awaitall},
  keywordstyle=\color{OliveGreen}\bfseries,
  keywords=[2]{T, i16, u16, i32, u32, i64, u64, f16, f32, readonly, writeonly, hops, channel, extern},
  keywordstyle=[2]\color{MidnightBlue}\bfseries,
  keywords=[3]{kernel, phase, place, compute, dataflow, if, else, while, for, foreach, map, async, in, return},
  keywordstyle=[3]\color{Mulberry}\bfseries,
  keywords=[4]{send, receive, @pipelined_reduce},
  keywordstyle=[4]\color{Bittersweet}\bfseries,
  identifierstyle=\color{identifier},
  sensitive=false,
  comment=[l]{//},
  morecomment=[s]{/*}{*/},
  commentstyle=\color{purple}\ttfamily,
  stringstyle=\color{red}\ttfamily,
  morestring=[b]',
  morestring=[b]"
}

\lstdefinelanguage{gt4py}{
  keywords={def},
  keywordstyle=\color{OliveGreen}\bfseries,
  keywords=[2]{Field3D, PARALLEL},
  keywordstyle=[2]\color{MidnightBlue}\bfseries,
  keywords=[3]{with},
  keywordstyle=[3]\color{Mulberry}\bfseries,
  keywords=[4]{@stencil, computation, interval, laplace},
  keywordstyle=[4]\color{Bittersweet}\bfseries,
  identifierstyle=\color{identifier},
  sensitive=false,
  comment=[l]{\#},
  commentstyle=\color{purple}\ttfamily,
  stringstyle=\color{red}\ttfamily,
  morestring=[b]',
  morestring=[b]"
}

\lstdefinelanguage{csl}{
  basicstyle=\fontencoding{T1}\selectfont\ttfamily\scriptsize,
  keywords={@activate,@unblock,@export,@export_symbol, @get_color, @get_dsd, @get_output_queue, @get_local_task_id, @mov16, @bind_local_task, @set_rectangle, @set_tile_code, @set_color_config},
 keywordstyle=\color{OliveGreen}\bfseries,
  keywords=[2]{f32, void, WEST, EAST, RAMP, fabout_dsd},
  keywordstyle=[2]\color{MidnightBlue}\bfseries,
  keywords=[3]{var, const, task, fn, for, layout, comptime},
  keywordstyle=[3]\color{Mulberry}\bfseries,
  keywords=[4]{sys_mod,unblock_cmd_stream},
  keywordstyle=[4]\color{Bittersweet},
  identifierstyle=\color{identifier},
  sensitive=false,
  comment=[l]{//},
  commentstyle=\color{purple}\ttfamily,
  stringstyle=\color{red}\ttfamily,
  morestring=[b]',
  morestring=[b]"
}

\crefname{listing}{listing}{listings}
\Crefname{listing}{Listing}{Listing}

\begin{document}

\author{
\IEEEauthorblockN{Lukas Gianinazzi$^{\star}$\thanks{$^{\star}$Equal contribution.}}
\IEEEauthorblockA{Noéda Research\\
Zurich, Switzerland\\
lux@noeda.ai}
\and
\IEEEauthorblockN{Tal Ben-Nun$^{\star}$}
\IEEEauthorblockA{Lawrence Livermore National Laboratory\\
Livermore, California, USA\\
}
\and
\IEEEauthorblockN{Torsten Hoefler}
\IEEEauthorblockA{Department of Computer Science\\
ETH Zurich\\
Zurich, Switzerland\\
}
}

\title{SpaDA: A Spatial Dataflow Architecture Programming Language}

\maketitle

\begin{abstract}
Spatial dataflow architectures like the Cerebras Wafer-Scale Engine deliver exceptional performance in AI and scientific computing by distributing scratchpad memory across hundreds of thousands of processing elements (PEs). Yet programming these architectures remains difficult: with no shared memory, data movement requires explicit configuration, and asynchronous task management introduces substantial complexity. We present SpaDA, a programming language that offers precise control over data placement, dataflow patterns, and asynchronous operations while abstracting low-level architectural details. We design and implement a compiler targeting Cerebras CSL through multi-level lowering and unique optimization passes. SpaDA functions as a high-level programming interface and an intermediate representation for domain-specific languages (DSLs), demonstrated here with the GT4Py stencil DSL. SpaDA enables concise expression of operations with complex parallel patterns -- including pipelined collective operations, multi-dimensional stencils, and dense linear algebra -- in 14.09$\times$ fewer lines than CSL, achieving over 260 TFlop/s across 730,000 PEs on a single device.
\end{abstract}
\maketitle

\section{Introduction}

Spatial dataflow architectures (SDAs) have emerged as high-performance platforms for memory-bound workloads, from neural network inference to stencil-based physical simulations in weather forecasting and computational fluid dynamics~\cite{DBLP:journals/corr/abs-2503-11698, DBLP:journals/micro/Lie24, DBLP:conf/sc/RockiESSMKPDS020, DBLP:conf/sc/SaiHMA24}. By distributing computation across hundreds of thousands of processing elements (PEs), each with fast local SRAM, SDAs eliminate the cache hierarchies and shared-memory contention of conventional architectures. In our work, we focus on the Cerebras Wafer-Scale Engine (WSE-2). It integrates over 730,000 PEs on a single die with 20 PB/s of peak memory bandwidth across 40 GB of SRAM~\cite{MiyajimaSC25}, 5000× higher peak memory bandwidth than an H100 GPU~\cite{nvidia_h100_datasheet_2022}. However, this remarkable performance comes at a steep programmability cost that has largely excluded the broader HPC community from exploiting it.

Programming the WSE is accomplished through the Cerebras Software Language (CSL), a low-level kernel language. CSL requires explicit orchestration of two hardware mechanisms with no analog in conventional parallel programming. First, PE communication flows through a circuit-switched network-on-chip where physical channels are scarce, finite resources. Concurrent data streams must be assigned distinct channels or routing conflicts lead to nondeterministic errors. Second, computation is driven asynchronously by data arrival and asynchronous tasks, requiring careful coordination across a distributed fabric with no shared state. Together, these constraints make achieving high hardware utilization challenging~\cite{DBLP:conf/asplos/StawinogaKLZ0BG26}.

Existing approaches offer no suitable general solution. Programming models developed for FPGAs and CGRAs emphasize loop scheduling such as tiling, unrolling, and pipelining~\cite{DBLP:conf/cgo/LichtKMBHH21}. Programs using MPI generally assume point-to-point communication capabilities~\cite{DBLP:conf/sc/LagunaMMRSS19}. All of these approaches assume packet-switched networks with dynamic routing and fundamentally cannot reason about the channel allocation and spatial placement decisions central to WSE programming.
Existing WSE applications in physical simulation~\cite{DBLP:conf/pldi/Castro-Perez0GY21, DBLP:conf/sc/JacquelinAM22, DBLP:conf/sc/SaiJHAS23} and machine learning~\cite{DBLP:journals/corr/abs-2304-03208, DBLP:journals/corr/abs-2112-07571},  rely on handcrafted CSL tightly coupled to specific algorithms~\cite{DBLP:conf/hpdc/LuczynskiGIWSH24,DBLP:conf/ics/Orenes-VeraSSJV23}. Common patterns like halo exchange must be reimplemented from scratch for each workload. 
The most recent stencil-focused DSLs~\cite{DBLP:conf/sc/SaiMXA24} and MLIR pipelines~\cite{DBLP:conf/asplos/StawinogaKLZ0BG26} for CSL narrow this gap for a single domain, but \emph{none provides a general-purpose programming model that generalizes across applications}.

\begin{figure}
    \centering
    \includegraphics[width=\linewidth]{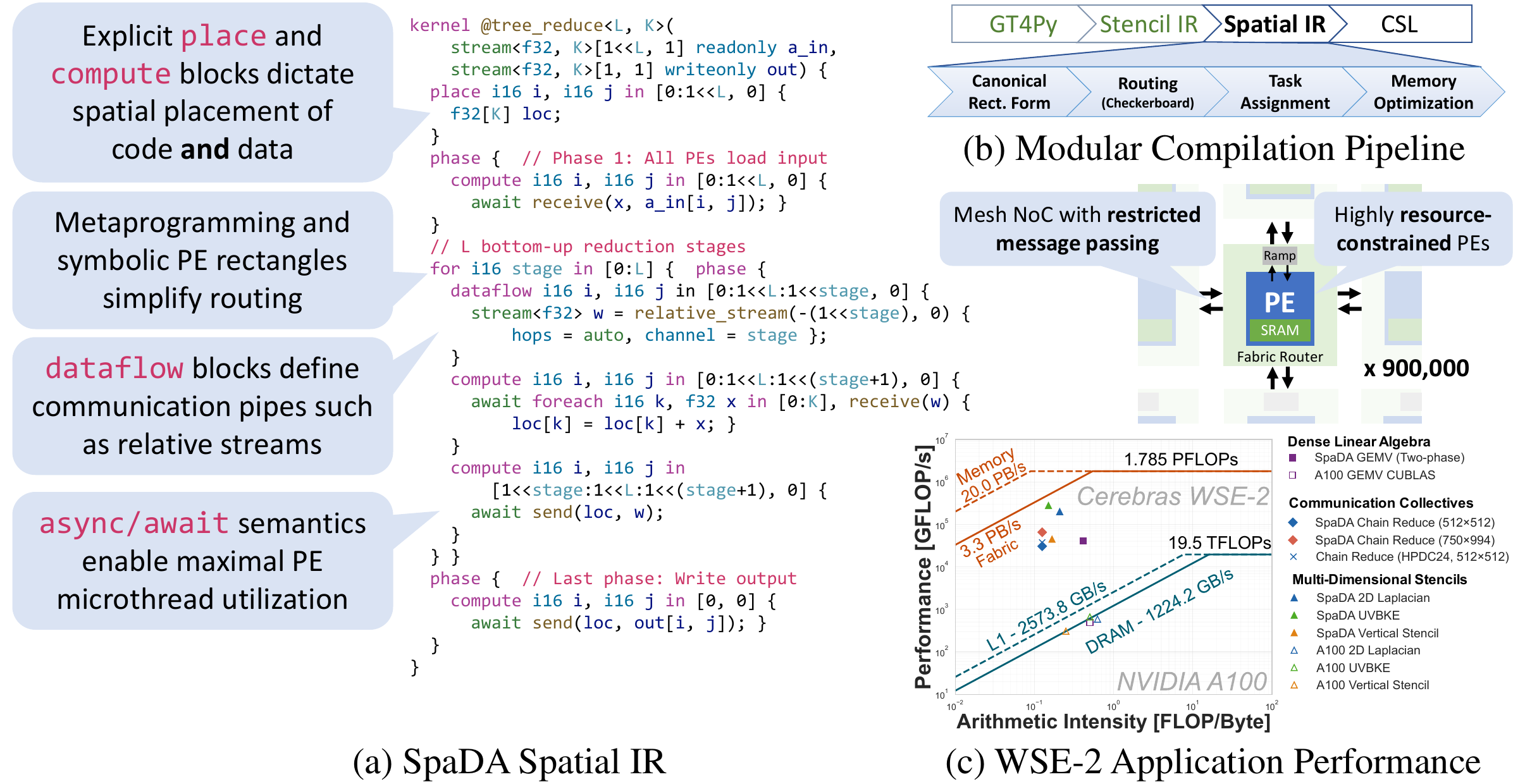}
    \caption{SpaDA Overview}
    \label{fig:SpaDA}
\end{figure}

We present \underline{Spa}tial \underline{D}ataflow \underline{A}bstraction (SpaDA), a programming language and optimizing compilation pipeline targeting CSL, that \textbf{makes the WSE programmable without sacrificing performance}. As shown in Figure~\ref{fig:SpaDA}a, SpaDA provides three constructs covering the full programming surface of the WSE: \emph{place} blocks for explicit data allocation across the PE grid, \emph{dataflow} blocks for declaring typed communication streams between PEs, and \emph{compute} blocks for asynchronous data-driven computation with async/await semantics. These abstractions give programmers precise control over data locality, communication structure, and computation-communication overlap, while hiding the low-level details of channel assignment, task scheduling, and vectorized computation that make raw CSL so difficult to use. SpaDA additionally serves as a compiler intermediate representation (IR) for domain-specific languages. As shown in Figure~\ref{fig:SpaDA}b, we demonstrate this through a complete modular compilation pipeline from GT4Py~\cite{DBLP:journals/corr/abs-2311-08322}, the production weather and climate stencil DSL running operational forecasts at CSCS and MeteoSwiss, through a Stencil IR and SpaDA's, to executable CSL. While SpaDA's language design is architecture-agnostic, our compiler backend and evaluation target the WSE specifically. The source code is available at \url{https://github.com/spcl/spada/}.
Our main contributions are:

\textbf{1. Language Design.} SpaDA provides high-level asynchronous constructs including async/await dependency management, foreach loops over data streams, and map scopes for vectorized operations. These enable concise expression of complex spatial algorithms spanning stencil computations, collective communication primitives, and dense linear algebra. SpaDA reduces code size by 4.68--13.13$\times$ compared to hand-written CSL for algorithmic kernels, and by up to 616$\times$ when lowering from GT4Py stencils.

\textbf{2. Optimizing Compilation Pipeline}
We present an optimizing compilation pipeline from GT4Py to SpaDA to CSL. We develop several novel optimization passes targeting CSL that leverage our expressive SpaDA representation. In particular, we implement an automatic routing assignment pass and graph-based task scheduling and fusion optimization passes. Moreover, we also implement a copy-elimination pass tailored to our asynchronous setting. 
In our results, we show that these passes are necessary to run several kernels at scale and overall improve task utilization by up to 2x and memory consumption by up to 50\%.

\textbf{3. Benchmarking} We evaluate SpaDA on a Cerebras WSE-2 across three categories of kernels: (1) Stencil Computations, (2) Communication Collectives, and (3) Dense Linear Algebra (GEMV).
We compare our results against previous work on WSE-2 and optimized A100 baselines. Compared to \textit{handwritten, near-optimal} reduction kernels, our generated code is only $1.04\times$ slower (harmonic mean). On stencils, we achieve up to 260 TFlops/s, which is over 400$\times$ higher throughput compared to the A100 baseline. On GEMV, we achieve over 80$\times$ higher throughput than the CUBLAS A100 baseline. Moreover, the WSE-2 results are more energy-efficient throughout. In particular, the SpaDA UVBKE stencil achieves 4.5$\times$ higher performance per Watt than the A100 baseline.


\section{Cerebras Wafer-Scale Engine and CSL}\label{sec:wse}

The Cerebras Wafer-Scale Engine (WSE) comprises 730,000--900,000 processing elements (PEs) arranged in a two-dimensional mesh topology. Each PE contains a SIMD-capable ALU and 48KB of local SRAM for storing data and code. Each PE accesses only its local memory with sub-nanosecond latency and can communicate with the other PEs by sending messages through the on-chip fabric~\cite{MiyajimaSC25}. 

\textbf{Circuit-Switched Communication.}
PEs communicate through a circuit-switched network-on-chip where data packets called \emph{wavelets} traverse configured routing paths. Each router supports a limited number (24 per PE and an additional 8 reserved) of concurrent circuits, \emph{colors} in CSL terminology, which are virtual communication channels bound to physical routing resources. 
Concurrent communication streams must be assigned distinct colors to avoid routing conflicts. 
As a result, when writing WSE applications color management quickly becomes a bottleneck, requiring judicious allocation.

\textbf{Task-Driven Execution.}
Computation in a WSE follows an asynchronous task model where \emph{local tasks} are activated programmatically and \emph{data tasks} trigger automatically upon wavelet arrival. A task runs only when both \emph{active} and \emph{unblocked}, with explicit \texttt{@activate} and \texttt{@unblock} operations enabling fine-grained synchronization. Task identifiers are also limited resources: up to 28 per PE. As data tasks are bound to colors, task IDs must not overlap with colors. Namely, using one task ID means that the same color ID is also blocked, reducing the effective local tasks and exacerbating the program scalability challenge.

\textbf{Vectorized Operations.}
High performance on the WSE ALU requires exploiting hardware-accelerated vectorized operations through \emph{Data Structure Descriptors} (DSDs). DSDs encode memory access patterns and fabric communication routes, enabling single-instruction operations like \texttt{@fadd}, \texttt{@fmul}, and \texttt{@fmac} to process entire arrays. Expressing computations as DSD operations rather than scalar loops is crucial for achieving peak PE utilization.

\textbf{CSL Code Structure.} 
The CSL syntax is based on the Zig language~\cite{zig}. CSL code is structured as a folder with code files that are loaded onto PEs, a layout CSL file that prescribes how code files and routing paths are mapped to the WSE, and a Python file that controls memory copy/streaming and kernel execution.
A minimal excerpt of a CSL PE code is as follows:
\begin{lstlisting}[language=csl,numbers=none]
// code_0.csl
var local_a: f32;  // Local variable
const out_dsd = @get_dsd(fabout_dsd, .{ .extent = 8, 
   .fabric_color = @get_color(1), 
   .output_queue = @get_output_queue(0) }); // Fabric DSD
   
task reset_a() void {
  local_a = 0.0;
  sys_mod.unblock_cmd_stream(); // Kernel is complete
}
fn entry_point() void {
  // DSD operation that transmits one to east neighbor
  @mov16(out_dsd, 1, .{.async=true,
                       .activate=@get_local_task_id(8)});
}
comptime {
  @bind_local_task(reset_a, @get_local_task_id(8));
  @export_symbol(entry_point);
}
\end{lstlisting}
and a corresponding layout file:
\begin{lstlisting}[language=csl,numbers=none]
// layout.csl
layout {
  @set_rectangle(2, 1);  // PE rectangle size
  // Computation setup
  @set_tile_code(0, 0, "code_0.csl");
  @set_tile_code(1, 0, "code_1.csl");
  // Communication setup (PE 0->1)
  @set_color_config(0, 0, @get_color(1), .{ 
    .routes = .{ .rx = .{RAMP}, .tx = .{EAST} } });
  @set_color_config(1, 0, @get_color(1), .{ 
    .routes = .{ .rx = .{WEST}, .tx = .{RAMP} } });
}
\end{lstlisting}

The code transmits the value 1 to the neighboring PE. A rectangle of size 2$\times$1 is setup in the layout file, each PE is assigned a separate code file, and communication is set up for each PE separately. Color assignment is restrictive; for instance, color 1 cannot be assigned a second time to communicate between PE (1,0) and (2,0). In \texttt{code\_0.csl}, we see an exported function (\texttt{fn}) and a \texttt{task}, bound to ID 8. The DSD operation \texttt{@mov16} will activate the task asynchronously after the message is sent. Communication is one-sided---no confirmation will be sent back upon receipt.

\subsection{Challenges in Authoring CSL Code}

CSL development has a substantially steeper learning curve than conventional kernel languages such as CUDA. It requires not only familiarity with one-sided communication, but also detailed knowledge of the hardware and its constraints. These constraints are evident even in vendor-provided examples. For instance, a simple 25-point stencil\footnote{https://sdk.cerebras.net/csl/code-examples/benchmark-25-pt-stencil} spans 887 lines of code and utilizes specialized hardware features.

A central difficulty is the scarcity of hardware resources. In particular, task IDs and colors, if assigned na\"{i}vely, will exhaust available resources quickly. In realistic CSL applications, developers often rely on low-level implementation patterns, including hand-coded state machines to preserve task IDs, tasks that block their own IDs, and manual DSD register management, such as changing base addresses to enable code reuse, and thus per-PE memory footprint reduction.

One-sided communication also makes correctness more difficult to achieve: if communication patterns across PE boundaries are not designed carefully, deadlocks can arise easily. The problem becomes even more pronounced for communication with higher-order neighbors (e.g. diagonally), necessitating the use of hardware-specific capabilities such as fabric router switches. 
Switches must be preprogrammed with a fixed number of routes controlled by wavelets,
and the burden of their management is solely on the developer.
%


Lastly, scalability at the program level is constrained by the limited per-PE memory available for both code and data, which makes larger applications difficult to implement. In practice, large CSL applications often require managing multiple CSL files for PE subgroups corresponding to boundary conditions, differing communication patterns, and specialized PE roles. When many such files are used, or when files are parameterized by PE location, compilation time and software management complexity increase significantly.

It is therefore imperative to provide a higher-level abstraction, which enables high utilization and maintainability.

\begin{table*}[htb]
\centering
\renewcommand{\arraystretch}{1.1}  
\caption{Core SpaDA Language Constructs}\label{tab:constructs}
\vspace{-0.75em}
\begin{tabular}{l l}
\toprule
\textbf{Construct} & \textbf{Description} \\
\midrule

|phase {...}| & Wraps a set of blocks to declare a phase. \\ 
\addlinespace[0.5em]
\multicolumn{2}{l}{Within a \textbf{\texttt{place}} block:} \\
|T scal|\hspace{0.5em}/\hspace{0.5em}|T[...] arr| & Declares a local scalar or (multidimensional) array of type \texttt{T} \\ 

\addlinespace[0.5em]
\multicolumn{2}{l}{Within a \textbf{\texttt{dataflow}} block:} \\
|stream<T> s = relative_stream(dx, dy)| & Declares a stream \texttt{s} of element type \texttt{T} \\ 
|stream<T> s = relative_stream([dx0:dx1], [dy0:dy1])| & Declares a multi-casting stream \texttt{s} of element type \texttt{T}\\

\addlinespace[0.5em]
\multicolumn{2}{l}{Within a \textbf{\texttt{compute}} block:} \\

|send(a, s)| & Asynchronously sends array \texttt{a} over stream \texttt{s} \\ 
|foreach f32 x in receive(s) {...}| & Iterates over each element received from the stream \texttt{s} \\
|foreach u16 k, f32 x in [0:J], receive(s) {...}| & Iterates over an element range received from stream \texttt{s} \\
|map i32 i in [I:J:K] {...}| & Asynchronous parallelizeable \emph{affine} loop \\
|async {...}| & Creates an asynchronous block of code \\ 
|completion c = ...| & Get a completion handle for an asynchronous operation  \\
|await c| & Waits for a specified completion \texttt{c} before proceeding with further computation \\ 
|awaitall| & Waits for all pending completions before proceeding \\
|for i64 i in [I:J:K] {...}| & Synchronous sequential loop \\ 


\bottomrule
\end{tabular}
\vspace{-1.25em}
\end{table*}

\section{SpaDA Language Design}\label{sec:language}

Our language framework enables precise control over data placement, data movement, and asynchronous execution, essential for programming spatial dataflow architectures effectively. This section introduces the main constructs: \texttt{place} blocks for data allocation across processing elements (PEs), \texttt{dataflow} blocks for setting up communication streams, and \texttt{compute} blocks for defining operations on both streamed and locally stored data. \Cref{tab:constructs} provides a brief reference of the language constructs available within these blocks.

Each block is defined for some \emph{subgrid}, which is specified using a range expression for each dimension.
%
%
In the example |place i32 i, i32 j in [0:I:2, 1:J-1] { ... }|, the variables \texttt{i} and \texttt{j} represent the coordinates of each PE in the subgrid |[0:I:2, 1:J-2]|, which covers every second PE in the first dimension (with a stride of 2) from $0$ to $I$ and every PE in the second dimension from $1$ to $J-2$ (exclusive).

Blocks can be organized into \emph{phases}, which serve as local temporal scopes. Streams and data allocated within a phase are only accessible within that specific phase. From the perspective of each processing element (PE), phases execute sequentially in the order they are defined in the code. However, phase transitions occur asynchronously across PEs, allowing one PE to move to the next phase even if other PEs are still completing the current one. To allow for more complex communication patterns, phases can be wrapped in one or more \textit{meta-programming} for-loop. The meta for-loop unrolls into a series of phases. For example, in \Cref{fig:SpaDA}a, each level of the binary communication tree corresponds to one iteration of this meta-programming for loop.

Together, these constructs provide a programming model that balances low-level control with high-level abstraction, allowing developers to write efficient, high-performance programs tailored to spatial architectures. We now examine each of the three block types in more detail.

\subsection{Place Blocks} 

The \texttt{place} block specifies where data is stored across the grid of processing elements (PEs). Within a \texttt{place} block, developers can allocate local data elements---either scalars or arrays---that are assigned to specific PEs based on their coordinates in a defined subgrid. Arrays and scalars contain undefined values upon initialization. 
Each \texttt{place} block defines memory allocations for the specified subgrid, forming a data layout across the processing grid. This layout ensures efficient data access for subsequent computations by placing data close to the PEs that will process it.

\subsection{Dataflow Blocks}
The \texttt{dataflow} block defines communication streams between PEs, establishing virtual data channels through which information flows. Each stream follows a relative communication pattern, meaning that connections are defined by offsets relative to each PE’s position in the subgrid. 
In particular, the declaration |stream<f32> s = relative_stream(x, y)| declares a stream that sends data from a PE at coordinate $(i, j)$ to a PE at coordinate $(i+x, j+y)$. If the stream is used for receiving, a PE at coordinate $(i, j)$ receives data from a PE at coordinate $(i-x, j-y)$. This implicit swapping of directionality when sending and receiving on the same stream enables pipelined communication chains (see \Cref{alg:1d-pipelined-reduce}).
Streams support \textit{multi-casting} in any single cardinal direction, allowing for efficient broadcasting of messages within subgrids.

Streams can be configured with specific routing paths and \texttt{channel} assignments, providing precise control over data movement. If two messages are routed through the same coordinate simultaneously, it is essential to ensure that they do not share the same \texttt{channel}. 

\subsection{Compute Blocks} 

The \texttt{compute} block defines the computations executed by each PE. Computations may be triggered by incoming data from streams, and can be asynchronous. This allows operations to overlap with data movement and thus minimizing idle time. Asynchronous constructs like \texttt{send}, \texttt{receive}, and \texttt{foreach} enable a data-driven model where computations adapt to the dynamic flow of information across the grid. 

To manage execution order and dependencies, \textit{completions} are used. Completions track the status of asynchronous tasks and ensure that each operation either has a unique completion assigned or is prefixed with \texttt{await}. The \texttt{await} operation synchronizes tasks by blocking until a specified completion is triggered or prefixed operation completes.
For instance, |await send(a, s)| ensures that the send operation finishes before moving to the next operation and |await c| for some completion \texttt{c} ensures that the operation associated with \texttt{c} has completed before the next operation starts.

Note that a send operation completes once the send buffer may be safely overwritten, not when the receiver has received the data.
Any completion not explicitly awaited is automatically awaited at the end of the \texttt{compute} block, guaranteeing that all tasks complete before the block exits.
We implicitly add an \texttt{awaitall} operation for all completions at the end of each \texttt{compute} block, locally synchronizing all pending operations before moving to the next phase.

This asynchronous model allows operations to be preempted and interleaved, maximizing parallelism and compute/communication overlap. By explicitly managing dependencies with |await|, data races can be avoided, ensuring predictable execution across PEs. The combination of |place|, |dataflow|, and |compute| blocks provides fine-grained control over data placement, movement, and processing on spatial architectures.

\subsection{Example: Pipelined Chain Reduce}

\begin{listing}[tb]
\begin{lstlisting}[numbers=none]
kernel @chain_reduce<K>(stream<f32>[K] readonly a_in,
                        stream<f32>[1] writeonly out) {
  place i16 i, i16 j in [0:K, 0] {
    f32[K] a
  } !\label{line:place_block}!
  // Phase 1: Read argument stream
  phase {
    compute i32 i, i32 j in [0:K, 0] {
      await receive(a, a_in[i])  !\label{line:data_loading}!
    }
  }
  // Phase 2: Perform reduction
  phase {
    // Declare dataflow
    dataflow i32 i, i32 j in [0:K, 0] {
      stream<f32> red = relative_stream(-1, 0)  !\label{line:stream_declare_red}!
      stream<f32> blue = relative_stream(-1, 0)  !\label{line:stream_declare_blue}!
    }
    // East corner
    compute i32 i, i32 j in [K-1, 0] {
      await send(a, red if (N-1) % 2 == 0 else blue) !\label{line:east_corner}!
    }
    // Odd PEs
    compute i32 i, i32 j in [1:K-1:2, 0] {
      await foreach i32 k, f32 x in [0:K], receive(red) {!\label{line:odd_receive}!
        a[k] = a[k] + x
        await send(a[k], blue)  !\label{line:odd_send}!
      }
    }
    // Even PEs
    compute i32 i, i32 j in [2:K-1:2, 0] {
      await foreach i32 k, f32 x in [0:K], receive(blue) {!\label{line:even_receive}!
        a[k] = a[k] + x
        await send(a[k], red)  !\label{line:even_send}!
      }
    }
    // West corner (root)
    compute i32 i, i32 j in [0, 0] {
      await foreach i32 k, f32 x in [0:K], receive(blue) {!\label{line:root_receive}!
        a[k] = a[k] + x
      }
      await send(a, out[i])  !\label{line:root_send}!
    }
  }
}
\end{lstlisting}
\vspace{-0.5em}
\caption{\footnotesize SpaDA 1D Pipelined `Chain' Reduction Kernel}
\label{alg:1d-pipelined-reduce}
\end{listing}
The kernel in Listing~\ref{alg:1d-pipelined-reduce} performs a 1D pipelined reduction across a row of processing elements (PEs), outputting the result at the west-most PE. This reduction is implemented in two phases: (1) data loading and (2) the pipelined reduction, followed by the output send.

The initial \texttt{place} block allocates a local array \texttt{a} of type |f32[K]| at each PE. 
In the data-loading phase, each PE asynchronously receives data from the input stream \texttt{a\_in}, into array \texttt{a}.
The reduction phase uses alternating \texttt{red} and \texttt{blue} streams. 

Using two streams ensures that no PE attempts to send and receive from the same stream simultaneously, which would otherwise cause communication conflicts and undefined behavior. The pipelined approach minimizes contention and is near-optimal for large vector sizes~\cite{DBLP:conf/hpdc/LuczynskiGIWSH24}.  This double-buffered strategy effectively maintains data flow without interruptions. 
In summary, this kernel demonstrates how our language’s constructs enable efficient pipelined reductions by providing explicit control over data movement and computation.

\subsection{Discussion} 

Our approach equips developers with precise control over dataflow, data placement, and asynchronous execution on spatial dataflow architectures. 
This programming model is particularly well-suited for applications requiring fine-grained control over data movement and processing, such as scientific computing, machine learning, and large-scale simulations. The explicit handling of data placement and communication enables fully exploiting the underlying hardware’s capabilities without low-level routing and synchronization, achieving both improved performance and productivity.

\section{Stencil DSL Frontend}\label{sec:dsl-frontend}

To further boost developer productivity, we demonstrate how SpaDA can serve as an intermediate representation for GT4Py~\cite{DBLP:journals/corr/abs-2311-08322,DBLP:conf/sc/BenNunGDWDDEGMTWFSH22} . 
GT4Py is a production embedded Python DSL used by CSCS and MeteoSwiss for weather forecasting. It allows domain scientists to express stencil computations using intuitive syntax. \Cref{alg:laplace-gt4py} shows a Laplacian in GT4Py:

\begin{listing}[tbh]
\begin{lstlisting}[language=gt4py,numbers=none]
@stencil
def laplace(in_field: Field3D, out_field: Field3D):
    with computation(PARALLEL), interval(...):
        out_field = -4.0 * in_field[0, 0, 0] + (
            in_field[1, 0, 0] + in_field[-1, 0, 0] + 
            in_field[0, 1, 0] + in_field[0, -1, 0])
\end{lstlisting}
\vspace{-0.5em}
\caption{\footnotesize Laplacian in GT4Py}
\label{alg:laplace-gt4py}
\end{listing}

This specification abstracts away data distribution, communication patterns, and synchronization. Direct lowering to SpaDA is challenging because GT4Py's domain-specific semantics (iteration intervals, vertical strategies, boundary conditions) don't map cleanly to SpaDA's explicit PE-level programming model.

We introduce a \emph{Stencil IR} that captures: (1) which field accesses require inter-PE communication versus local computation, (2) what halo regions boundary PEs need to satisfy neighbor dependencies, and (3) the data types and iteration domains for all operations. This intermediate representation decouples high-level stencil semantics from spatial code generation, enabling reuse across multiple stencil frontends and architecture-independent optimizations.

The Stencil IR is lowered to SpaDA through three simultaneous passes. The \emph{placement pass} allocates local arrays on each PE based on computed field sizes and halos. The \emph{dataflow pass} identifies communication patterns from stencil access offsets and generates stream declarations---for the Laplacian, the four neighbor accesses at offsets $(\pm1, 0, 0)$ and $(0, \pm1, 0)$ become four |relative_stream| declarations. The \emph{compute pass} transforms each stencil statement into SpaDA operations, inserting |send|/|receive| pairs where neighbor data crosses PE boundaries and generating |map| operations for local computation. Rectangle splitting and merging algorithms coalesce operations with identical subgrids, reducing code duplication.

\section{CSL Compiler Backend}\label{sec:compiler}

Lowering SpaDA to executable Cerebras CSL code requires bridging several fundamental abstraction gaps: SpaDA expresses computations over logical PE grids with abstract communication streams, while CSL demands explicit hardware task graphs, data structure descriptors (DSDs), and physical routing configurations. This transformation must preserve SpaDA's asynchronous semantics while exploiting CSL's hardware-accelerated primitives: vectorized DSD operations, wavelet-triggered tasks, and circuit-switched fabric communications.

We implement a multi-stage compilation pipeline that systematically lowers SpaDA's high-level constructs to CSL's architecture-specific features. 
%
%
(1) Canonicalization normalizes SpaDA into standard form with unique files for each PE type. (2) Routing assignment guarantees conflict-free channel allocation through checkerboard decomposition, producing routed SpaDA. (3) Task assignment extracts completion DAGs and coarsens them into CSL tasks, while adhering to resource constraints. (4) Code generation synthesizes DSD-optimized CSL code with pattern matching. (5) Memory optimization and I/O mapping configure CSL memory layouts, communication modes, and reduce PE memory. We now detail each stage.

\subsection{Canonicalization}

The lowering process begins with passes that normalize SpaDA into a form amenable to hardware mapping. These passes: (a) consolidate rectangles into \textit{PE equivalence classes} mapped to non-overlapping strided regions, ensuring each PE corresponds to a single CSL code file without generating a code file per PE; (b) unify |phase| blocks with |awaitall| synchronization markers, standardizing each subgrid to contain exactly one |place|, |dataflow|, and |compute| block; and (c) decompose high-level array operations on fields into explicit |foreach| or |map| blocks with index calculations.
%

\subsection{Routing Optimization}\label{sec:routing-assignment}

\begin{figure}[t]
    \centering
    \includegraphics[width=\linewidth,trim={3.2cm 5cm 6cm 3cm},clip]{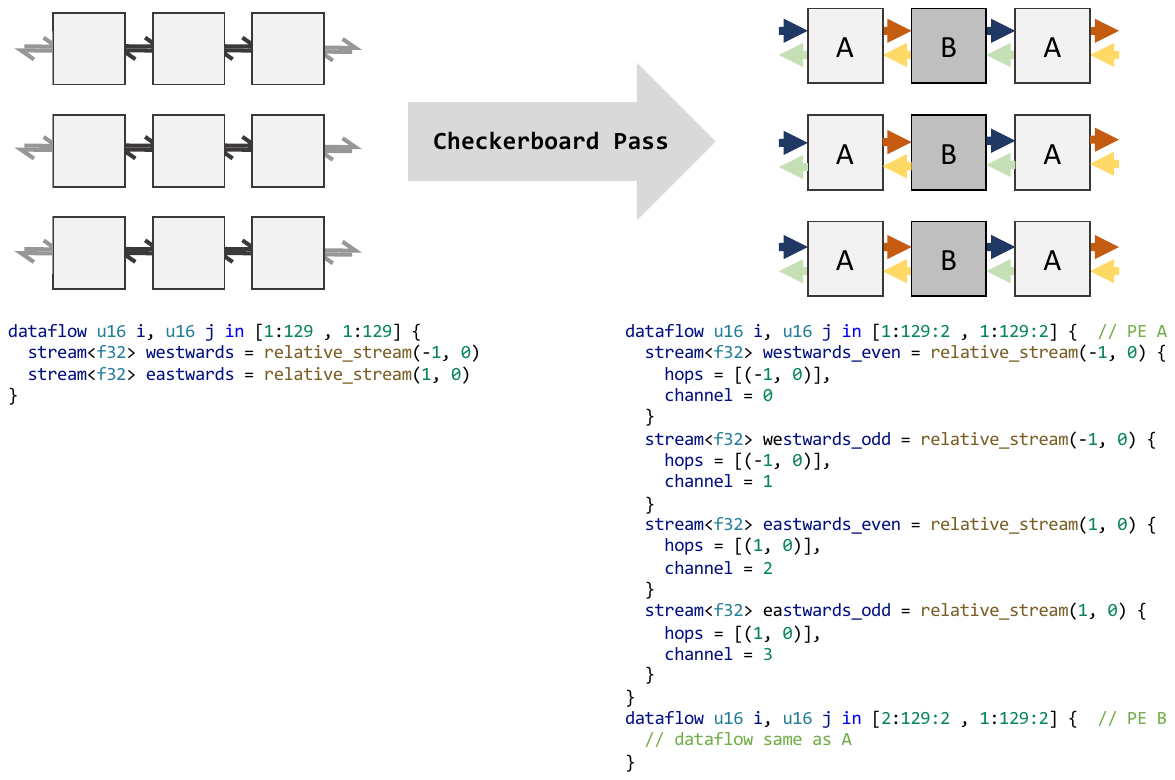}
    \vspace{-1.5em}
    \caption{Checkerboard Decomposition Pass (One Dimension)}
    \label{fig:checkerboard}
\end{figure}

The routing assignment stage addresses a challenge unique to spatial dataflow architectures: \emph{automatic channel allocation with conflict avoidance}. When multiple data streams traverse shared physical routing resources, they must be assigned to distinct channels to prevent race conditions and undefined behavior. We solve this through a checkerboard decomposition algorithm (\cref{fig:checkerboard}) that guarantees conflict-free routing by construction. 

\textbf{Checkerboard Decomposition.}
We first identify \emph{active dimensions}: a dimension is active if any stream has non-zero offset in that dimension. Each |compute| block is then split based on PE coordinate parity into 2 blocks for each active dimension. Each stream |s| is duplicated into |s_even| and |s_odd|. The key insight is that messages from even-coordinate PEs travel only through even-coordinate intermediate PEs, while messages from odd-coordinate PEs traverse only odd-coordinate PEs. This eliminates all routing conflicts.

At each |send| or |receive| operation on stream |s = relative_stream(dx, dy)|, we replace |s| with either |s_even| or |s_odd| depending on the parity and communication direction.
Our current checkerboard implementation restricts streams to single-hop communication ($
\|dx\| + \|dy\| \leq 1$), which suffices for all weather stencils in our evaluation. The algorithm can be extended to deal with multi-hop scenarios. Manually allocated channels already support multi-hop communication.

\textbf{Layout and Resource Allocation.}
The compiler also generates layout specifications that map SpaDA's abstract communication channels to physical CSL colors. A global allocation algorithm analyzes all subgrids to assign colors based on the aforementioned conflict-avoiding routing assignment. Subsequently, the compiler generates per-subgrid routing configurations with directional fabric paths, and input/output queues are assigned to the PEs' fabric DSDs to improve concurrency. 

\subsection{Task Graph Optimization}

\begin{figure}[t]
    \centering
    \includegraphics[width=\linewidth,trim={0cm 1.2cm 0.25cm 0cm},clip]{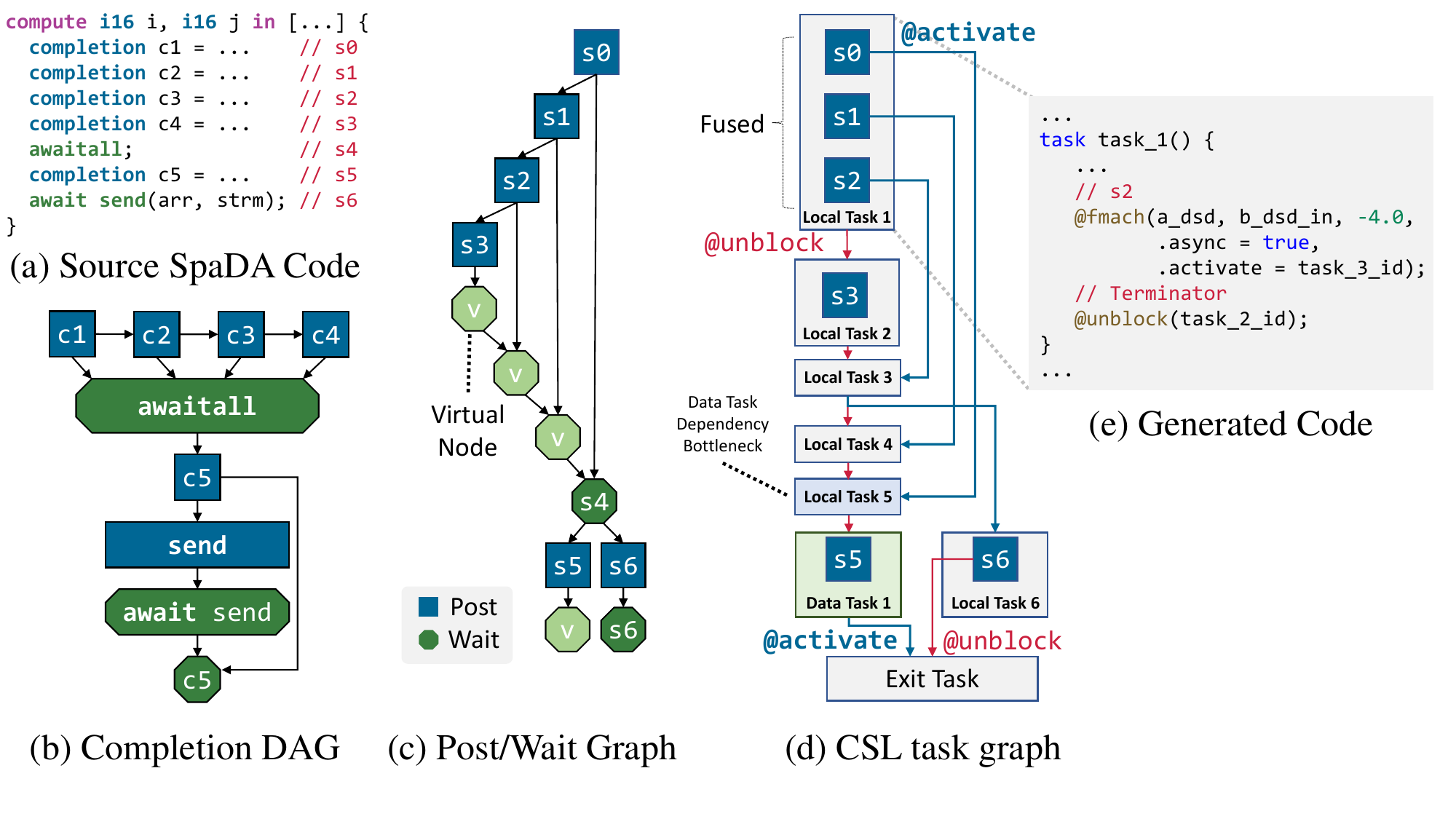}
    \caption{SpaDA-to-CSL Task Assignment Pipeline}
    \label{fig:tasks}
\end{figure}

In addition to color assignment, a key limited resource to utilize when creating WSE programs is tasks. Tasks can execute concurrently on a PE and the asynchrony is required to avoid deadlocks on nonblocking communication. Since SpaDA is an async-await language, we transform the code into a task graph in multiple stages, as outlined in  \cref{fig:tasks}. This representation enables applying optimizations on the task graph, namely \emph{fusion} and \emph{task ID recycling} (i.e., virtualization).

First, the compute block is transformed into a dependency DAG based on |completion| and |await| statements (\Cref{fig:tasks}b). This completion DAG is then converted to a scheduling graph where every statement is explicit, containing ``post'' and ``wait'' events (\Cref{fig:tasks}c), where synchronous statements are implemented as a post-wait sequence. The low-level statement decomposition allows us to later group them into local and data tasks according to a semantics-based rule-set (e.g., whether the statement can be implemented by a DSD operation).

Owing to the WSE PE architecture (\Cref{sec:wse}), task execution is constrained.
In particular, a CSL program must adhere to the following constraints: (a) a local task may have up to two predecessors (triggering the task with \texttt{@activate} and \texttt{@unblock}); and (b) a data task is always active, but can be blocked (thus having one predecessor). We write a set of passes to create \textit{virtual} nodes and local tasks to reduce the in-degree of nodes in the post/wait graph accordingly.

In the next step (\Cref{fig:tasks}d), the post/wait graph is coarsened to tasks based on feasibility, and edges from statements to successor tasks are annotated with \texttt{activate}/\texttt{unblock}. This allows the compiler to embed the activation into an asynchronous DSD operation (see \Cref{fig:tasks}e for an example), where a special \texttt{terminator} virtual statement can create dependencies at the end of a task.

We implement two optimization passes, which are crucial for compiling larger programs. These aim to reduce the number of used task IDs, which compete together with the router channels on the WSE. In particular, the tree-reduce communication collective would not compile without both of these optimizations (See \Cref{sec:ablations}).

\textbf{Task Fusion} is a coarsening optimization pass that groups statements into CSL tasks when possible, reducing overhead and the number of generated tasks.

\textbf{Task Recycling} allows mapping multiple logical tasks to a single hardware task ID. We generate a dispatch task that uses a state machine to decide which logical task to execute. Correctness relies on guaranteeing that two logical tasks that have the same hardware task ID may never run concurrently, and moreover, the predecessors of the later task may not run concurrently with the earlier task. 
We build a task conflict graph, then use a greedy load-balancing coloring~\cite{Besta20COLOR} to both reduce the number of task IDs and the number of logical tasks with the same task ID.

In summary, these passes allow the user to use async/await constructs to manage concurrency, while the compiler ensures task resources are adequately scheduled and managed.

\subsection{Automatic Vectorization}

SpaDA's \texttt{foreach} and \texttt{map} constructs are particularly amenable to automatic vectorization through DSDs. The compiler employs pattern matching to identify loops that correspond to DSD operations based on type and content (e.g., \texttt{@fmac*} for fused multiply-accumulate, and \texttt{@mov*} for data movement). 
When vectorization fails, the compiler employs a tiered fallback strategy. Loop bodies satisfying purity constraints (single output, indexing-only iterator usage, no control flow) are transformed into CSL \texttt{@map} operations, where the loop body becomes a callback function with variables promoted to parameters and output assignments converted to returns. Alternatively, \texttt{foreach} loops over streams without explicit ranges become wavelet-triggered data tasks, where fabric colors trigger execution upon packet arrival. Complex control flow, non-affine indexing, or other non-conforming patterns fall back to direct CSL for-loop generation. This hierarchical approach maximizes hardware utilization through DSD operations while maintaining semantic correctness through conservative fallbacks.

\subsection{Memory Optimization.}

\textbf{I/O Mapping.}
SpaDA kernel arguments do not explicitly occur in |place| blocks. Therefore, a mapping pass reserves memory for copy results on the PE called |extern| fields. To determine the local size and the mapping between streams and corresponding PEs, an analysis pass inspects the generated task graph and array slice expressions. 
Scalar arguments are implemented in CSL as kernel function arguments that are |place|d only in PEs that use them.

\textbf{Copy Elimination.}
SpaDA |place| blocks may contain short-lived intermediate fields that serve as staging buffers between adjacent operations and input/output streams. Such intermediates directly compete with application data for the scarce 48KB local memory per PE. The copy-elimination pass reclaims this overhead by identifying fields with a single producer and a single consumer, or |extern| fields with only consumers or producers (for input and output arguments respectively), substituting the source directly at the use site, and pruning unused fields. It operates at two granularities: whole-field forwarding across flat statement regions and indexed forwarding inside loop bodies.

The SpaDA CSL runtime uses the aforementioned I/O metadata and the compiled CSL to execute the kernel code.

\section{Results}\label{sec:experiments}

We run the experiments on a Cerebras WSE-2 system using CSL SDK v1.4.0. The WSE-2 chip comprises a fabric of $757\times 996$ PEs (out of which $750\times 994$ are usable due to the on-chip memcpy infrastructure). We measure each kernel 100 times and collect cycle counters from all participating PEs. The numbers reported here are median over the maximal cycle count among all PEs (to account for imbalanced workloads), with 95\% nonparametric CI for error bars. If 95\% CI is under 0.01, it is omitted for readability. For easier interpretability, we convert the number of cycles into a runtime with the formula:
$
\text{Runtime [}\mu\text{s]} = (\text{cycles} / 0.85) \cdot 10^{-3}
$.

Our evaluation comprises three categories of kernels. 
First, we examine communication collectives. 
These serve as fundamental building blocks for distributed linear algebra operations and are crucial for machine learning training. 
We based our designs on the algorithms from Luczynski et al.~\cite{DBLP:conf/hpdc/LuczynskiGIWSH24}, who presented near-optimal communication collectives for the WSE. Namely, \emph{Chain Reduce} (See \Cref{alg:1d-pipelined-reduce}), \emph{Tree Reduce} (See \Cref{fig:SpaDA}), and a hybrid approach called \emph{Two-Phase Reduce}.  
We compare our SpaDA implementations against Luczynski et al.'s handwritten CSL results. 
Note that for the broadcast benchmark, we synchronize the clocks of the PEs (with an approach similar to Luczynski et al.) to obtain accurate results.

Second, we evaluate three 3D stencil kernels, originally written in GT4Py and lowered through our DSL compilation pipeline (\Cref{sec:dsl-frontend}): a 2D Laplacian computing spatial derivatives in the horizontal plane, a difference stencil with sequential dependencies along the vertical column direction, and UVBKE, a complex momentum equation kernel from the COSMO dynamical weather core with horizontal computations. 
All stencil experiments use $K=80$ vertical (pressure) levels unless otherwise specified.
We compare our results with an NVIDIA A100 40GB GPU, using GT4Py version 1.1.7.

Third, we use our SpaDA communication collectives to implement a GEMV $\bm{y} = \alpha \bm{A}\bm{x} + \beta \bm{y}$. We implement a 1.5D partitioned $\bm{A}$-stationary algorithm~\cite{DBLP:conf/ics/SelvitopiBNTYB21}. 
This algorithm partitions $\bm{A}$ into equal-sized blocks and accordingly also partitions $\bm{x}$ equally among the columns of the grid and $\bm{y}$ among the rows of the grid. The algorithm broadcasts $\bm{x}$ from north to south, then performs the local matrix-vector products, then reduces the partial results from west to east. Finally, the result is computed in the first row of the subgrid locally. We compare two SpaDA variants using either Chain Reduce or Two Phase Reduce. Our baseline is CUBLAS 13.1.1 on an NVIDIA A100 40GB GPU.

Note that both the WSE-2 and the A100 use a 7nm manufacturing process.

\subsection{Developer Productivity}

\Cref{tab:loc-comparison} compares code size across representations. SpaDA kernels require 4.68--13.13$\times$ fewer lines than equivalent hand-written CSL. For GT4Py stencils, expansion patterns reveal spatial complexity: the vertical stencil shows modest growth due to its simple sequential structure, while horizontal stencils expand dramatically: the 2D Laplacian grows from 4 lines to nearly 2500 lines of CSL. This reflects the extensive PE coordination and halo exchange logic required for distributed spatial computation.

\begin{table}[tbh]
\centering
\caption{Lines of Code Across Representations}
\label{tab:loc-comparison}
\begin{tabular}{lcccc}
\toprule
\textbf{Kernel} & \textbf{GT4Py} & \textbf{SpaDA} & \textbf{CSL} & \textbf{CSL/Source} \\
\midrule
1D Broadcast & --- & 23 & 302$^{\dag}$ &13.13$\times$ \\  
2D Chain Reduction & --- & 91 & 606$^{\dag}$ & 6.66$\times$\\  
2D Tree Reduction & --- & 54 & 689$^{\dag}$ &12.76$\times$ \\  
2D Two-Phase Reduction & --- & 146 & 683$^{\dag}$ &4.68$\times$ \\  
Vertical Stencil & 5 & 26 & 53 & 10.6$\times$ \\
2D Laplacian & 4 & 1600 & 2464 & 616$\times$ \\  
UVBKE & 10 & 1222 & 2085 & 208.5$\times$ \\ 
GEMV & --- & 88 & 4586 & 52.11$\times$ \\  
GEMV Two-Phase & --- & 122 & 472572 & 3873.54$\times$ \\  
\midrule 
\textbf{Harmonic Mean} & --- & --- & --- & 14.09$\times$ \\  
\bottomrule
\end{tabular}\vspace{0.25em}
${^\dag}$ handwritten kernel~\cite{DBLP:conf/hpdc/LuczynskiGIWSH24}
\vspace{-0.5em}
\end{table}

As mentioned in \Cref{sec:language}, meta-programming significantly simplifies writing tree reductions, producing higher lines of code reductions (12.76$\times$) over other more complex schemes. Complex communication rules that generate specialization, such as Two-Phase GEMV, tend to generate more layout code (where 419,994 lines reside in the layout file).

These results validate SpaDA's dual role. As a programming language, it significantly reduces hand-written code  while maintaining explicit control over performance-critical decisions. As a compiler IR, it enables tractable DSL lowering by separating domain semantics (\Cref{sec:dsl-frontend}) from architecture details (\Cref{sec:compiler}).
In summary, SpaDA provides an effective abstraction boundary for productive SDA programming.

\subsection{Communication Collectives}

\begin{figure}[t]
    \centering
    \includegraphics[width=.85\linewidth]{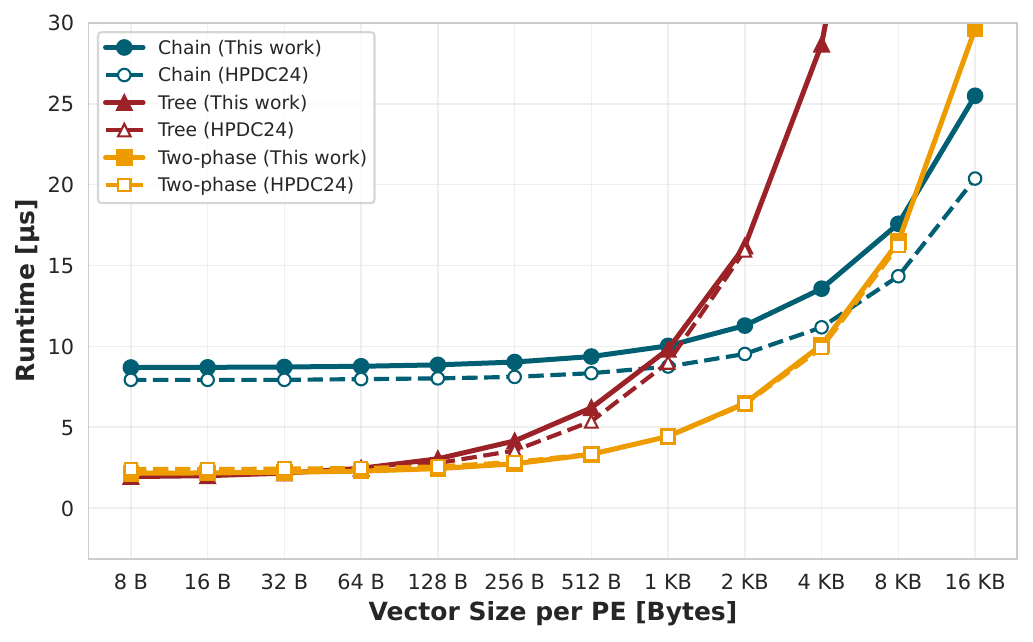}
    \vspace{-1em}
    \caption{2D Reduce Collectives (512$\times$512 $\approx$ 262K PEs)}
    \label{fig:collective-sweep}
\end{figure}

\Cref{fig:collective-sweep} compares SpaDA written reduce collective kernels with the results reported by Lucznyski et al.~\cite{DBLP:conf/hpdc/LuczynskiGIWSH24} (HPDC24), including the near-optimal Two-Phase reduction proposed in that work. We can see that our implementations are only $1.04\times$ slower compared to the hand-written CSL (harmonic mean across the entire message spectrum). The SpaDA versions closely follow the same tradeoffs as the baseline. 

\begin{figure}[t]
    \centering
    \includegraphics[width=.85\linewidth]{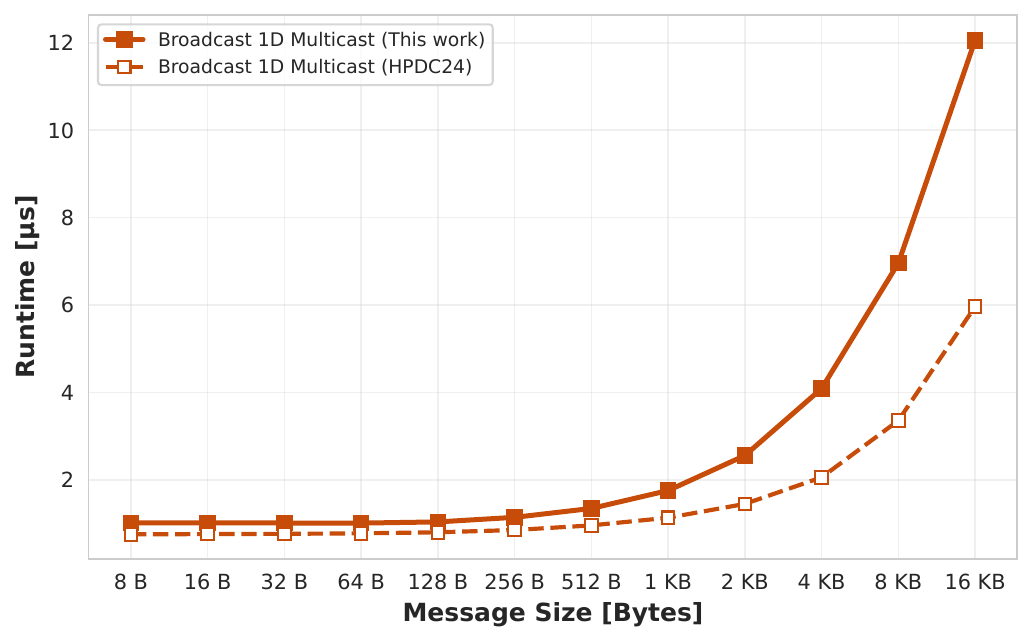}
    \vspace{-0.5em}
    \caption{1D Broadcast Collectives (512$\times$1 PEs) 
    }
    \label{fig:broadcast-sweep}
\end{figure}

\Cref{fig:broadcast-sweep} compares a SpaDA broadcast kernel with the results from Luczynski et al.~\cite{DBLP:conf/hpdc/LuczynskiGIWSH24}. Our broadcast kernel has 30\%-100\% overhead compared to the handwritten kernel. As we use the optimal number of DSD operations (one), we attribute this to differences in the hardware, SDK version (1.4 vs 1.0), or measurement setup.

\subsection{Stencil Computations}

\begin{figure}[t]
    \centering
    \includegraphics[width=.85\linewidth]{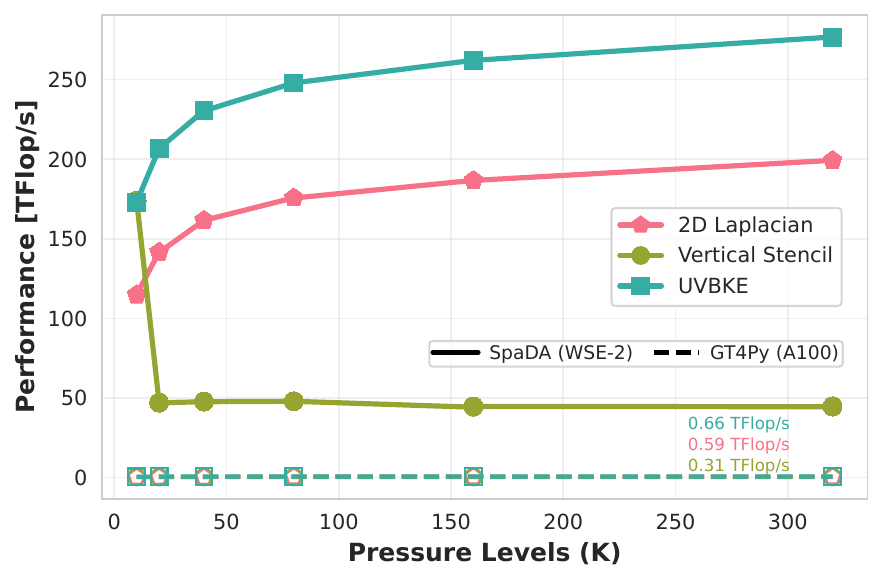}
    \vspace{-0.5em}
    \caption{Stencil FLOP/s for Fixed Horizontal Domain (746$\times$990) on $\approx$ 730K PEs and Varying Vertical Domain Size}
    \label{fig:stencil-sweep-vertical}
\end{figure}

Figure~\ref{fig:stencil-sweep-vertical} examines performance when fixing the horizontal domain while varying the number of vertical levels. The horizontal stencils, Laplacian and UVBKE, scale well with vertical levels because each level represents independent parallel work. UVBKE achieves over $260$ TFLOP/s. 

In contrast, the vertical stencil exhibits different behavior: throughput increases up to 17 pressure levels, then drops significantly. 
This is because up until 16 pressure levels, the CSL compiler unrolls the sequential loop.
The vertical stencil's sequential dependencies along each vertical column execute within a single PE, preventing parallelization across levels. 

\subsection{Linear Algebra}

\begin{figure}[t]       
    \centering
    \includegraphics[width=.85\linewidth]{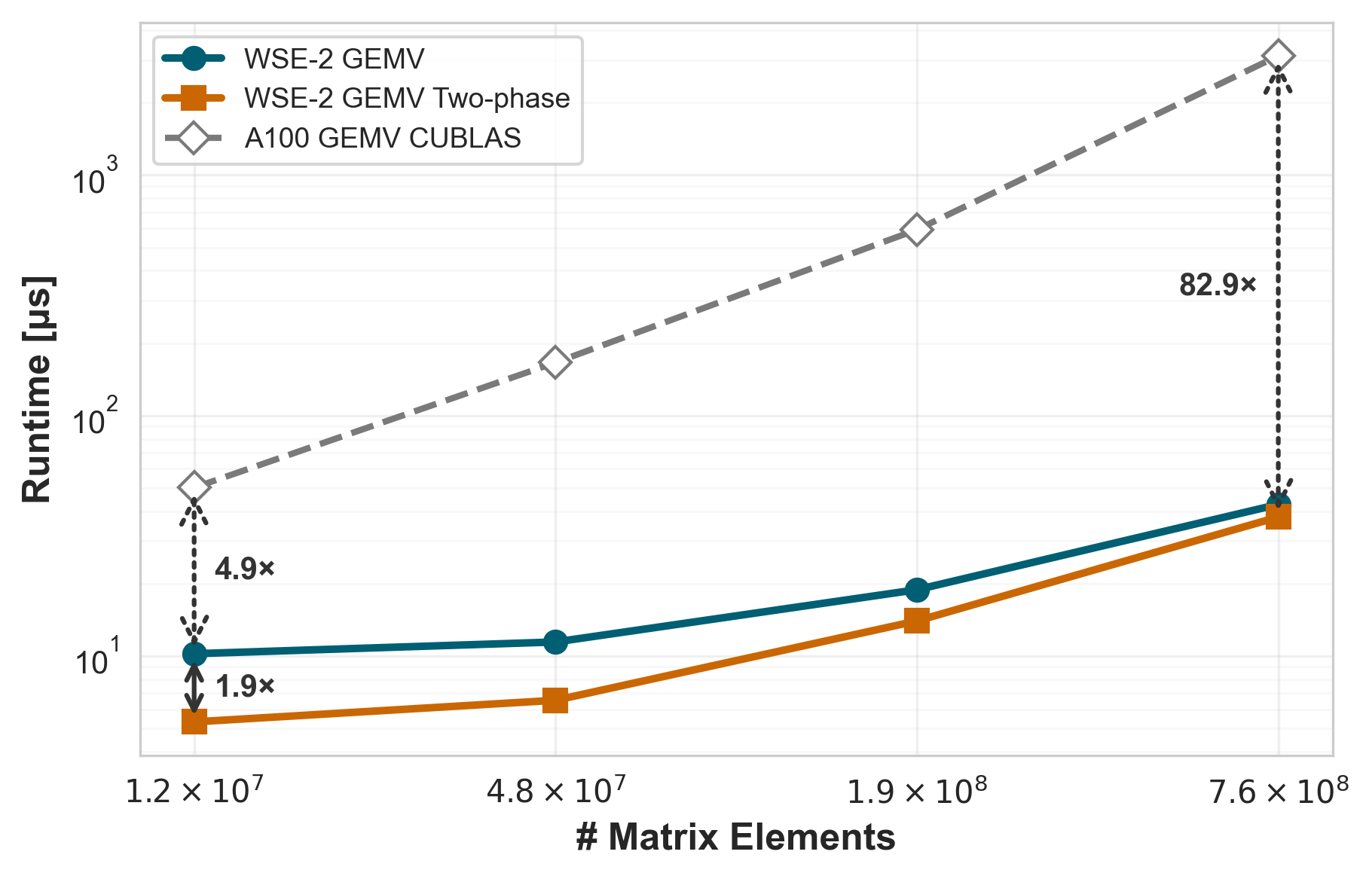}
    \vspace{-0.5em}
    \caption{GEMV Runtime for Varying Matrix Size $\bm{A}$}
    \label{fig:gemv-scaling}
\end{figure}

Compared with CUBLAS on A100, our SpaDA implementations achieved a speedup of up to 82.9$\times$. \Cref{fig:gemv-scaling} shows that the GEMV using two-phase reduction is up to 1.9$\times$ faster than the version using the chain reduce. This is because the number of elements transferred is relatively small compared to the overall memory used, since for each $k \times k$ block of the matrix $\bm{A}$, only $2k$--$3k$ elements are transferred.

We also launched the 1D partitioned GEMV benchmark from the Cerebras SDK\footnote{\href{https://github.com/Cerebras/sdk-examples/tree/rel-sdk-1.4.0/benchmarks/gemv-collectives\_2d}{https://github.com/Cerebras/sdk-examples/tree/rel-sdk-1.4.0/benchmarks/gemv-collectives\_2d}}, but it ran OOM for all matrix sizes larger than 2048$\times$2048, due to the SDK benchmark's distribution scheme, which does not partition the inputs $\bm{y}$ and $\bm{x}$. For this matrix size, Cerebras's code ran for 15{,}410 cycles, which is \textbf{5.46$\times$} slower than our two-phase 1.5D partitioned GEMV (at 2{,}822 cycles), as well as our direct GEMV implementation (5{,}597 cycles).




\subsection{Peak Performance}

We provide a roofline model to provide insights into the utilisation of the systems. 
We use the same roofline parameters as Jacquelin et al.~\cite{DBLP:conf/sc/JacquelinAM22}. 
Note that while the WSE-2 theoretical peak SRAM memory bandwidth is 20 PB/s, STREAM benchmarks observed an effective memory bandwidth of 8.8 PB/s~\cite{MiyajimaSC25}. 
The theoretical peak bandwidth from the fabric to the PE (`off/on-ramp to fabric') is 3.3 PB/s. Note that the theoretical full duplex fabric bandwidth is 27.5 PB/s, but, following Jacquelin et al., we do not consider it as one of the rooflines as it is never the limiting factor in practice. 
We count both accesses to the fabric and local memory.

\begin{figure}[t]
    \centering
    \includegraphics[width=.9\linewidth]{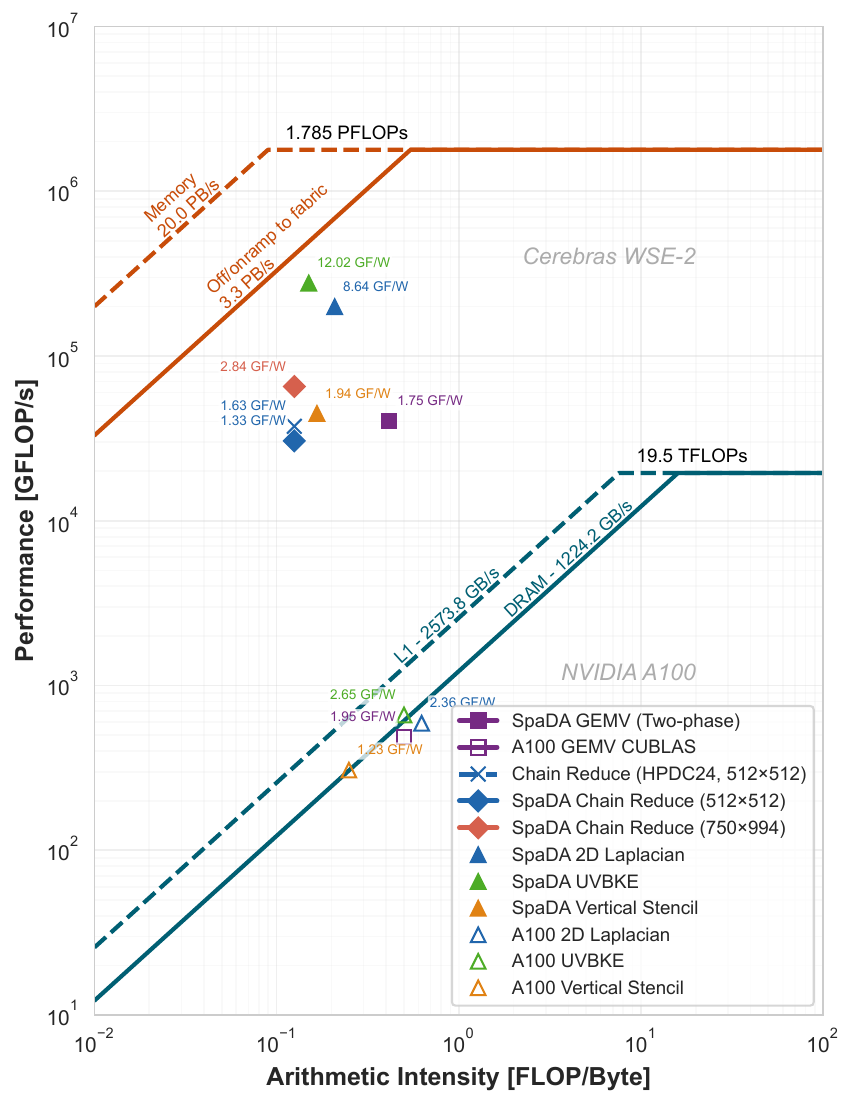}
    \vspace{-0.5em}
    \caption{Roofline plot comparing SpaDA on WSE-2 against A100 baselines and previous work. Performance per Watt is annotated in GFLOP/s/Watt (GF/W).}
    \label{fig:roofline}
\end{figure}

The roofline plot in \cref{fig:roofline} reveals that all kernels except GEMV are bound by the fabric-to-PE bandwidth. For the UVBKE and 2D Laplacian stencils, we can see that we are close to the off/onramp to fabric bandwidth. 
For GEMV, there is still significant potential for improving the PE-local matrix-vector multiply. We used a na\"{i}ve dot-product formulation, which might explain the distance from the roofline.

The A100 kernels are highly optimized and hit the DRAM bandwidth.
Nevertheless, they exhibit two orders of magnitude lower throughput compared to our SpaDA implementations.


\subsection{Power Efficiency}

The WSE-2 operates at a reported 16.5kW~\cite{DBLP:conf/sc/LtaiefHWJRK23} -- 23kW~\cite{DBLP:conf/sc/JacquelinAM22}. 
The A100 has a peak power draw of 250 W~\cite{nvidia_a100_datasheet_2020}. \Cref{fig:roofline} shows that the estimated performance per Watt of SpaDA versus the A100 baselines is generally favorable, with our SpaDA stencils achieving up to 12 GFlop/s/Watt. 

\subsection{Compiler Pass Ablation Study}\label{sec:ablations}

\begin{figure}[t]
    \centering
    \subfloat[UVBKE (domain size: 746$\times$990$\times$320)]{
    \includegraphics[width=\linewidth,clip,trim={0cm 0.1cm 0cm 0cm}]{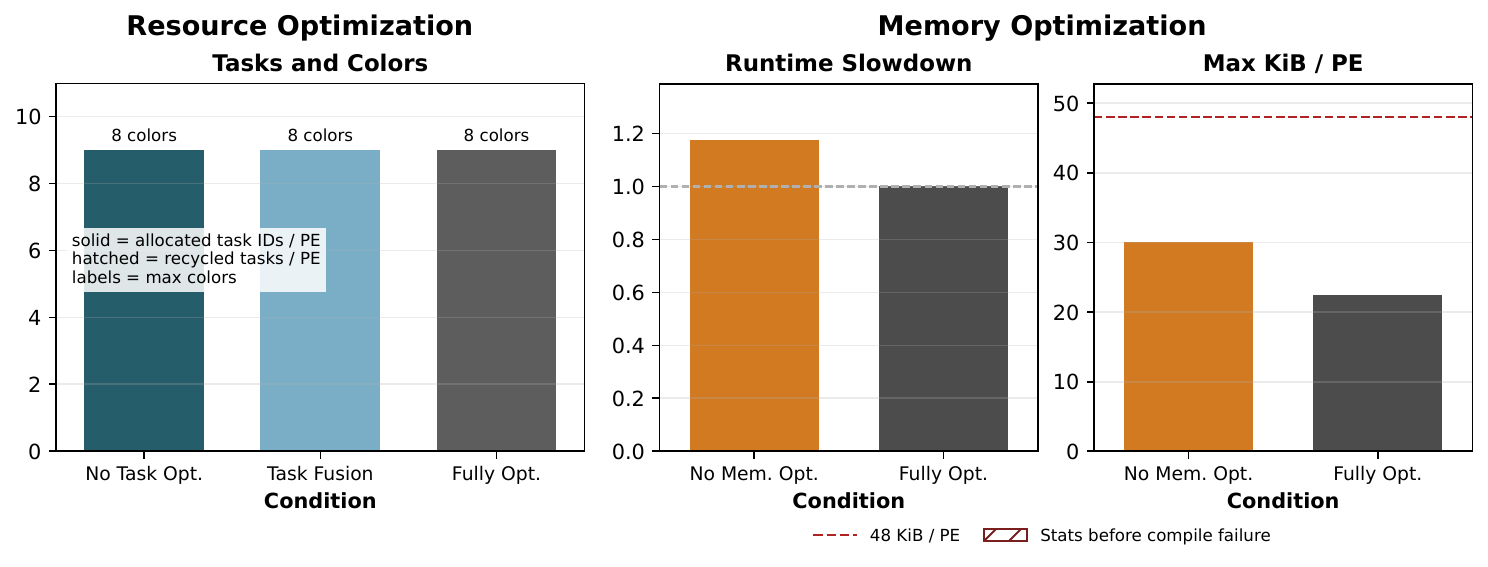}\label{fig:ablation:uvbke}}\\\vspace{-0.5em}
    \subfloat[Tree 2D Reduce (512$\times$512 PEs, 1 KB)]{\includegraphics[width=\linewidth,clip,trim={0cm 0.1cm 0cm 0cm}]{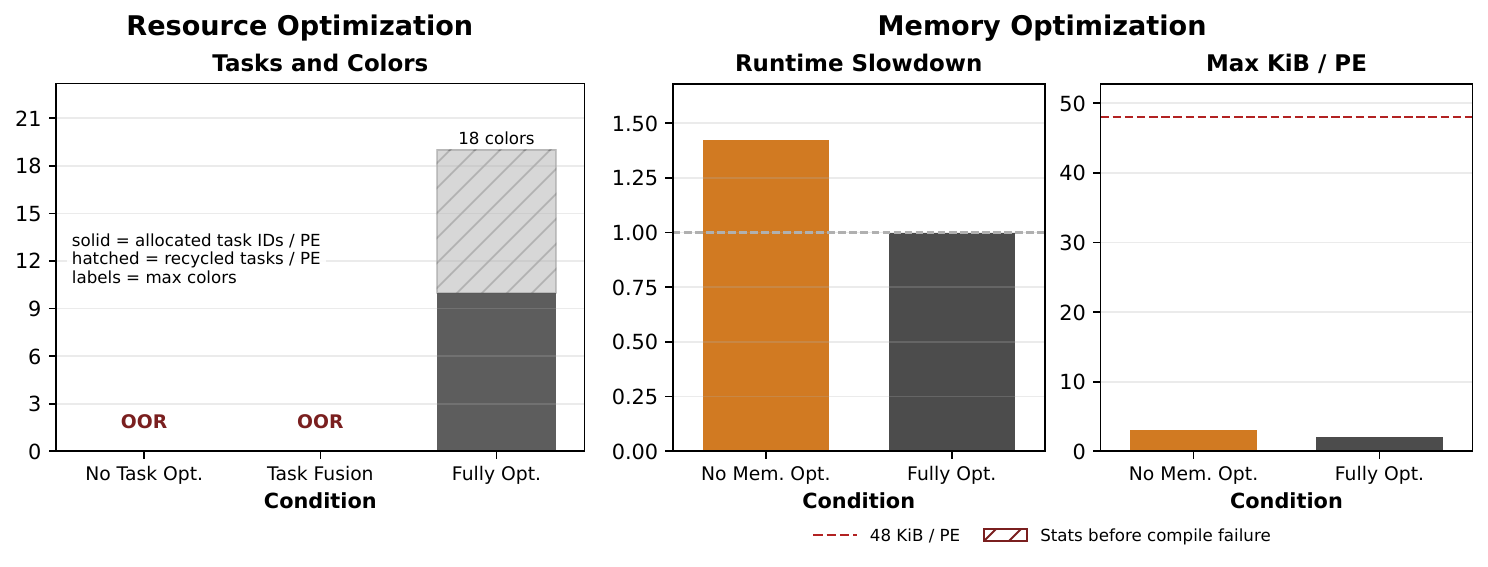}\label{fig:ablation:tree}}\\\vspace{-0.5em}
    \subfloat[Two-Phase 2D Reduce (512$\times$512 PEs, 16 KB)]{\includegraphics[width=\linewidth,clip,trim={0cm 0.1cm 0cm 0cm}]{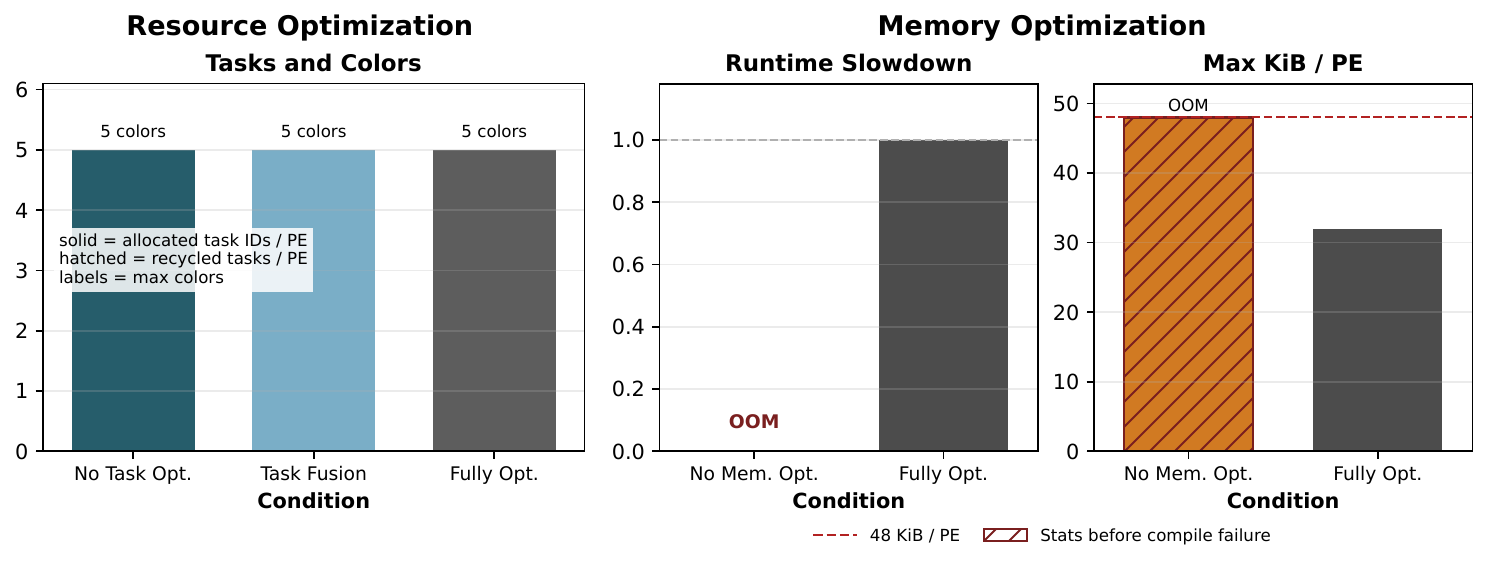}\label{fig:ablation:twophase}}
    \caption{Compiler optimization ablation study depicting performance and PE resource utilization when task fusion, task recycling, and copy elision optimizations are disabled. OOR=Out of Resources (e.g., colors, tasks), OOM=Out of Memory.}
    \label{fig:ablation}
\end{figure}

We present ablation studies for three crucial optimization passes, namely task fusion, task ID recycling, and copy elimination. \Cref{fig:ablation} showcases the optimizations and their effect on performance and resource allocation.

For kernels with simple communication patterns and low resident memory requirements, such as the UVBKE stencil (\cref{fig:ablation:uvbke}), memory optimization improves performance and reduces memory footprint. The latter has the benefit of enabling larger stencil graphs to be mapped onto the device. For collectives with more complex communication patterns, such as the two-phase and tree two-dimensional reduction, we see that even primitives such as collectives can easily over-allocate resources on the device, requiring the state machine due to increased use of both colors and asynchronous tasks. Tree reduction in particular (\cref{fig:ablation:tree}), consumes $2\log_2{P}$ colors (one for each dimension times the tree depth). This makes scheduling impossible without carefully managing task allocation, which SpaDA provides to developers effortlessly. For cases such as two-phase reduction (\cref{fig:ablation:twophase}) and others (e.g., broadcast with multicast), not performing memory optimization can limit the effective local field size that can be used.

In all cases, the passes also contribute to the end-to-end performance by reducing the number of DSD operations and task scheduling overhead.

\subsection{Discussion}

Our results demonstrate that SpaDA significantly reduces programming complexity for spatial dataflow architectures. Developers can express complex patterns in tens of lines while achieving predictable performance by maximizing parallelism and arithmetic intensity. 
Several optimization opportunities remain. Our checkerboard decomposition guarantees conflict-free routing but the number of colors could be further optimized. 
Moreover, our experiments revealed optimization opportunities for compute-bound kernels. 

\section{Related Work}

\subsection{Optimizing Stencil Operations}

The Open Earth Compiler~\cite{DBLP:journals/taco/GysiMZHDWFHG21} demonstrates multi-level IR rewriting from high-level stencil dialects through MLIR to GPU code. While sharing our focus on progressive lowering, they target GPU architectures with shared memory and loop transformations, whereas SpaDA addresses spatial dataflow architectures with distributed memory and circuit-switched routing.
StencilFlow~\cite{DBLP:conf/cgo/LichtKMBHH21} maps stencil DAGs to distributed FPGA systems with automated deadlock analysis. Like SpaDA, it targets spatial architectures, but differs fundamentally in targeting reconfigurable hardware with packet-switched communication rather than hardened dataflow processors requiring explicit channel assignment.

\subsection{Computations on the Cerebras WSE}

Prior work on the Cerebras WSE has demonstrated impressive performance through hand-crafted implementations~\cite{DBLP:conf/pldi/Castro-Perez0GY21, DBLP:conf/sc/JacquelinAM22, DBLP:conf/sc/SaiJHAS23, DBLP:conf/sc/RockiESSMKPDS020, DBLP:conf/sc/SaiHMA24, SantosSC24, BrownEuroPar23, DBLP:conf/sc/LtaiefHWJRK23}, but each required extensive manual effort tailored to specific algorithms. Communication strategies, memory layouts, and computational kernels were tightly coupled to a particular solution strategy, and common patterns had to be reimplemented for each application.

Luczynski et al.~\cite{DBLP:conf/hpdc/LuczynskiGIWSH24} achieved speedups for hand-optimized reductions, requiring manual reasoning about routing conflicts across hundreds of thousands of PEs---patterns SpaDA can express naturally with automatic conflict-free routing.

The WSE Field-equation API (WFA)~\cite{DBLP:journals/corr/abs-2209-13768} improved programmability with a NumPy-like interface but remains domain-specific to stencil computations on uniform Cartesian grids with fixed domain decomposition. 
Sai et al.~\cite{DBLP:conf/sc/SaiMXA24} presented their own stencil DSL and CSL compilation pipeline.
Most recently, Stawinoga et al. presented a MLIR lowering pipeline for stencils to CSL
~\cite{DBLP:conf/asplos/StawinogaKLZ0BG26}.
SpaDA differs in scope and design from these stencil-focused approaches: SpaDA can express diverse parallel patterns beyond stencils, including pipelined reductions with alternating communication and collective primitives and dense linear algebra, and serves dual roles as both a programming language and compiler IR.
This way, SpaDA offers the convenience of a DSL (through our GT4Py frontend) and the flexibility of a kernel programming language without the burden of low-level CSL.

\subsection{Programming Frameworks}
Our work draws on research in session types, static analysis for message-passing programs, and formal semantics for concurrent systems, while addressing unique challenges posed by spatial dataflow architectures.

\emph{Multiparty session types}~\cite{DBLP:conf/popl/HondaYC08,DBLP:conf/icalp/DenielouY13,DBLP:conf/pldi/Castro-Perez0GY21} enable communication safety through global-to-local projection but assume packet-switched networks with transparent routing. SpaDA extends this framework with explicit routing paths and hardware channel assignment for circuit-switched NoCs with finite physical channels. 
\emph{Pabble}~\cite{DBLP:conf/pdp/NgY14} introduces parameterized session types for MPI collective operations but lacks spatial placement and routing primitives. 
\emph{Bronevetsky}'s communication-sensitive dataflow~\cite{DBLP:conf/cgo/Bronevetsky09} infers communication patterns from existing MPI code through parallel control-flow graphs. SpaDA is prescriptive: explicit specification of topology, placement, and routing enables automatic routing assignment and guaranteed safety by construction---reflecting that spatial architectures require explicit data movement orchestration. 

\emph{GPU Tile Programming} Recently, Tilus~\cite{DBLP:conf/asplos/DingHZL0Y0P26} presented an approach to describe the layout of tensors onto tiles using layout functions. While Tilus addresses a related problem of data placement on GPUs, its scope is inherently different as it leaves out the spatial routing assignment inherent to SDAs.

\section{Conclusion}

SpaDA demonstrates that spatial dataflow architectures can be programmed productively without sacrificing performance. By separating data placement, communication, and computation from hardware-specific routing and task management, complex parallel algorithms become expressible concisely and without performance loss.
%
%
SpaDA provides a foundation for future work on compiler optimization passes and lowers the barrier for porting new applications to WSE.

\section*{Acknowledgments}
Work by Lawrence Livermore National Laboratory was performed under the auspices of the U.S. Department of Energy under contract DE-AC52-07NA27344 (LLNL-CONF-2012380).

Generative AI was used to improve the language and overall presentation of this paper~\cite{gpt54,ClaudeAI}.

\bibliographystyle{IEEEtran}
\bibliography{bibliography} 

@inproceedings{DBLP:conf/asplos/StawinogaKLZ0BG26,
  author       = {Nicolai Stawinoga and
                  David Katz and
                  Anton Lydike and
                  Justs Zarins and
                  Nick Brown and
                  George Bisbas and
                  Tobias Grosser},
  editor       = {Benjamin C. Lee and
                  Harry Xu and
                  Mark Silberstein and
                  Bingyao Li},
  title        = {An {MLIR} Lowering Pipeline for Stencils at Wafer-Scale},
  booktitle    = {Proceedings of the 31st {ACM} International Conference on Architectural
                  Support for Programming Languages and Operating Systems, Volume 2,
                  {ASPLOS} 2026, Pittsburgh, PA, USA, March 22-26, 2026},
  pages        = {94--109},
  publisher    = {{ACM}},
  year         = {2026},
  url          = {https://doi.org/10.1145/3779212.3790124},
  doi          = {10.1145/3779212.3790124},
  bibsource    = {dblp computer science bibliography, https://dblp.org}
}

@article{DBLP:journals/corr/abs-2209-13768,
  author       = {Mino Woo and
                  Terry Jordan and
                  Robert Schreiber and
                  Ilya Sharapov and
                  Shaheer Muhammad and
                  Abhishek Koneru and
                  Michael James and
                  Dirk Van Essendelft},
  title        = {Disruptive Changes in Field Equation Modeling: {A} Simple Interface
                  for Wafer Scale Engines},
  journal      = {CoRR},
  volume       = {abs/2209.13768},
  year         = {2022},
  url          = {https://doi.org/10.48550/arXiv.2209.13768},
  doi          = {10.48550/ARXIV.2209.13768},
  eprinttype    = {arXiv},
  eprint       = {2209.13768},
  bibsource    = {dblp computer science bibliography, https://dblp.org}
}

@inproceedings{DBLP:conf/ics/Orenes-VeraSSJV23,
  author       = {Marcelo Orenes{-}Vera and
                  Ilya Sharapov and
                  Robert Schreiber and
                  Mathias Jacquelin and
                  Philippe Vandermersch and
                  Sharan Chetlur},
  editor       = {Kyle A. Gallivan and
                  Efstratios Gallopoulos and
                  Dimitrios S. Nikolopoulos and
                  Ram{\'{o}}n Beivide},
  title        = {Wafer-Scale Fast Fourier Transforms},
  booktitle    = {Proceedings of the 37th International Conference on Supercomputing,
                  {ICS} 2023, Orlando, FL, USA, June 21-23, 2023},
  pages        = {180--191},
  publisher    = {{ACM}},
  year         = {2023},
  url          = {https://doi.org/10.1145/3577193.3593708},
  doi          = {10.1145/3577193.3593708},
  bibsource    = {dblp computer science bibliography, https://dblp.org}
}

@article{DBLP:journals/corr/abs-2112-07571,
  author       = {Meredith V. Trotter and
                  Cuong Q. Nguyen and
                  Stephen Young and
                  Rob T. Woodruff and
                  Kim M. Branson},
  title        = {Epigenomic language models powered by Cerebras},
  journal      = {CoRR},
  volume       = {abs/2112.07571},
  year         = {2021},
  doi           = {10.48550/arXiv.2112.07571},
  url          = {https://doi.org/10.48550/arXiv.2112.07571},
  eprinttype    = {arXiv},
  eprint       = {2112.07571},
  bibsource    = {dblp computer science bibliography, https://dblp.org}
}

@article{DBLP:journals/micro/Lie24,
  author       = {Sean Lie},
  title        = {Inside the Cerebras Wafer-Scale Cluster},
  journal      = {{IEEE} Micro},
  volume       = {44},
  number       = {3},
  pages        = {49--57},
  year         = {2024},
  url          = {https://doi.org/10.1109/MM.2024.3386628},
  doi          = {10.1109/MM.2024.3386628},
  bibsource    = {dblp computer science bibliography, https://dblp.org}
}

@article{DBLP:journals/corr/abs-2503-11698,
  author       = {Yudhishthira Kundu and
                  Manroop Kaur and
                  Tripty Wig and
                  Kriti Kumar and
                  Pushpanjali Kumari and
                  Vivek Puri and
                  Manish Arora},
  title        = {A Comparison of the Cerebras Wafer-Scale Integration Technology with
                  Nvidia GPU-based Systems for Artificial Intelligence},
  journal      = {CoRR},
  volume       = {abs/2503.11698},
  year         = {2025},
  url          = {https://doi.org/10.48550/arXiv.2503.11698},
  doi          = {10.48550/ARXIV.2503.11698},
  eprinttype    = {arXiv},
  eprint       = {2503.11698},
  bibsource    = {dblp computer science bibliography, https://dblp.org}
}

@article{DBLP:journals/corr/abs-2304-03208,
  author       = {Nolan Dey and
                  Gurpreet Gosal and
                  Zhiming Chen and
                  Hemant Khachane and
                  William Marshall and
                  Ribhu Pathria and
                  Marvin Tom and
                  Joel Hestness},
  title        = {Cerebras-GPT: Open Compute-Optimal Language Models Trained on the
                  Cerebras Wafer-Scale Cluster},
  journal      = {CoRR},
  volume       = {abs/2304.03208},
  year         = {2023},
  url          = {https://doi.org/10.48550/arXiv.2304.03208},
  doi          = {10.48550/ARXIV.2304.03208},
  eprinttype    = {arXiv},
  eprint       = {2304.03208},
  bibsource    = {dblp computer science bibliography, https://dblp.org}
}

@inproceedings{DBLP:conf/hpdc/LuczynskiGIWSH24,
  author       = {Piotr Luczynski and
                  Lukas Gianinazzi and
                  Patrick Iff and
                  Leighton Wilson and
                  Daniele De Sensi and
                  Torsten Hoefler},
  editor       = {Patrizio Dazzi and
                  Gabriele Mencagli and
                  David K. Lowenthal and
                  Rosa M. Badia},
  title        = {Near-Optimal Wafer-Scale Reduce},
  booktitle    = {Proceedings of the 33rd International Symposium on High-Performance
                  Parallel and Distributed Computing, {HPDC} 2024, Pisa, Italy, June
                  3-7, 2024},
  pages        = {334--347},
  publisher    = {{ACM}},
  year         = {2024},
  url          = {https://doi.org/10.1145/3625549.3658693},
  doi          = {10.1145/3625549.3658693},
  bibsource    = {dblp computer science bibliography, https://dblp.org}
}

@inproceedings{DBLP:conf/sc/SaiJHAS23,
  author       = {Ryuichi Sai and
                  Mathias Jacquelin and
                  Fran{\c{c}}ois P. Hamon and
                  Mauricio Araya{-}Polo and
                  Randolph R. Settgast},
  title        = {Massively Distributed Finite-Volume Flux Computation},
  booktitle    = {Proceedings of the {SC} '23 Workshops of The International Conference
                  on High Performance Computing, Network, Storage, and Analysis, {SC-W}
                  2023, Denver, CO, USA, November 12-17, 2023},
  pages        = {1713--1720},
  publisher    = {{ACM}},
  year         = {2023},
  url          = {https://doi.org/10.1145/3624062.3624252},
  doi          = {10.1145/3624062.3624252},
  bibsource    = {dblp computer science bibliography, https://dblp.org}
}

@inproceedings{DBLP:conf/sc/LtaiefHWJRK23,
  author       = {Hatem Ltaief and
                  Yuxi Hong and
                  Leighton Wilson and
                  Mathias Jacquelin and
                  Matteo Ravasi and
                  David Elliot Keyes},
  editor       = {Dorian Arnold and
                  Rosa M. Badia and
                  Kathryn M. Mohror},
  title        = {Scaling the "Memory Wall" for Multi-Dimensional Seismic Processing
                  with Algebraic Compression on Cerebras {CS-2} Systems},
  booktitle    = {Proceedings of the International Conference for High Performance Computing,
                  Networking, Storage and Analysis, {SC} 2023, Denver, CO, USA, November
                  12-17, 2023},
  pages        = {6:1--6:12},
  publisher    = {{ACM}},
  year         = {2023},
  url          = {https://doi.org/10.1145/3581784.3627042},
  doi          = {10.1145/3581784.3627042},
  bibsource    = {dblp computer science bibliography, https://dblp.org}
}

@InProceedings{BrownEuroPar23,
author="Brown, Nick
and Echols, Brandon
and Zarins, Justs
and Grosser, Tobias",
editor="Singer, Jeremy
and Elkhatib, Yehia
and Blanco Heras, Dora
and Diehl, Patrick
and Brown, Nick
and Ilic, Aleksandar",
title="Exploring the Suitability of the Cerebras Wafer Scale Engine for Stencil-Based Computation Codes",
booktitle="Euro-Par 2022: Parallel Processing Workshops",
year="2023",
publisher="Springer Nature Switzerland",
address="Cham",
pages="51--65",
isbn="978-3-031-31209-0",
url={https://doi.org/10.1007/978-3-031-31209-0_4},
doi={10.1007/978-3-031-31209-0_4}
}

@inproceedings{SantosSC24,
  author       = {Kylee Santos and
                  Stan G. Moore and
                  Tomas Oppelstrup and
                  Amirali Sharifian and
                  Ilya Sharapov and
                  Aidan P. Thompson and
                  Delyan Z. Kalchev and
                  Danny Perez and
                  Robert Schreiber and
                  Scott Pakin and
                  Edgar A. Leon and
                  James H. Laros III and
                  Michael James and
                  Sivasankaran Rajamanickam},
  title        = {Breaking the Molecular Dynamics Timescale Barrier Using a Wafer-Scale
                  System},
  booktitle    = {Proceedings of the International Conference for High Performance Computing,
                  Networking, Storage, and Analysis, {SC} 2024, Atlanta, GA, USA, November
                  17-22, 2024},
  eid          = {8},
  publisher    = {{IEEE}},
  year         = {2024},
  url          = {https://doi.org/10.1109/SC41406.2024.00014},
  doi          = {10.1109/SC41406.2024.00014},
  bibsource    = {dblp computer science bibliography, https://dblp.org}
}

@inproceedings{DBLP:conf/sc/SaiMXA24,
  author       = {Ryuichi Sai and
                  John M. Mellor{-}Crummey and
                  Jinfan Xu and
                  Mauricio Araya{-}Polo},
  title        = {Automated Code Generation of High-Order Stencils for a Dataflow Architecture},
  booktitle    = {Proceedings of the International Conference for High Performance Computing,
                  Networking, Storage, and Analysis, {SC} 2024, Atlanta, GA, USA, November
                  17-22, 2024},
  eid          = {19},
  publisher    = {{IEEE}},
  year         = {2024},
  url          = {https://doi.org/10.1109/SC41406.2024.00025},
  doi          = {10.1109/SC41406.2024.00025},
  bibsource    = {dblp computer science bibliography, https://dblp.org}
}

@inproceedings{DBLP:conf/sc/SaiHMA24,
  author       = {Ryuichi Sai and
                  Fran{\c{c}}ois P. Hamon and
                  John M. Mellor{-}Crummey and
                  Mauricio Araya{-}Polo},
  title        = {Matrix-Free Finite Volume Kernels on a Dataflow Architecture},
  booktitle    = {Proceedings of the International Conference for High Performance Computing,
                  Networking, Storage, and Analysis, {SC} 2024, Atlanta, GA, USA, November
                  17-22, 2024},
  eid          = {28},
  publisher    = {{IEEE}},
  year         = {2024},
  url          = {https://doi.org/10.1109/SC41406.2024.00034},
  doi          = {10.1109/SC41406.2024.00034},
  bibsource    = {dblp computer science bibliography, https://dblp.org}
}

@inproceedings{DBLP:conf/sc/BenNunGDWDDEGMTWFSH22,
  author       = {Tal Ben{-}Nun and
                  Linus Groner and
                  Florian Deconinck and
                  Tobias Wicky and
                  Eddie Davis and
                  Johann Dahm and
                  Oliver Elbert and
                  Rhea George and
                  Jeremy McGibbon and
                  Lukas Tr{\"{u}}mper and
                  Elynn Wu and
                  Oliver Fuhrer and
                  Thomas C. Schulthess and
                  Torsten Hoefler},
  editor       = {Felix Wolf and
                  Sameer Shende and
                  Candace Culhane and
                  Sadaf R. Alam and
                  Heike Jagode},
  title        = {Productive Performance Engineering for Weather and Climate Modeling
                  with Python},
  booktitle    = {{SC22:} International Conference for High Performance Computing, Networking,
                  Storage and Analysis, Dallas, TX, USA, November 13-18, 2022},
  pages        = {73:1--73:14},
  publisher    = {{IEEE}},
  year         = {2022},
  url          = {https://doi.org/10.1109/SC41404.2022.00078},
  doi          = {10.1109/SC41404.2022.00078},
  bibsource    = {dblp computer science bibliography, https://dblp.org}
}

@article{DBLP:journals/corr/abs-2311-08322,
  author       = {Enrique G. Paredes and
                  Linus Groner and
                  Stefano Ubbiali and
                  Hannes Vogt and
                  Alberto Madonna and
                  Kean Mariotti and
                  Felipe A. Cruz and
                  Lucas Benedicic and
                  Mauro Bianco and
                  Joost VandeVondele and
                  Thomas C. Schulthess},
  title        = {GT4Py: High Performance Stencils for Weather and Climate Applications
                  using Python},
  journal      = {CoRR},
  volume       = {abs/2311.08322},
  year         = {2023},
  url          = {https://doi.org/10.48550/arXiv.2311.08322},
  doi          = {10.48550/ARXIV.2311.08322},
  eprinttype    = {arXiv},
  eprint       = {2311.08322},
  bibsource    = {dblp computer science bibliography, https://dblp.org}
}

@inproceedings{Besta20COLOR,
author = {Besta, Maciej and Carigiet, Armon and Janda, Kacper and Vonarburg-Shmaria, Zur and Gianinazzi, Lukas and Hoefler, Torsten},
title = {High-performance parallel graph coloring with strong guarantees on work, depth, and quality},
year = {2020},
isbn = {9781728199986},
publisher = {IEEE Press},
booktitle = {Proceedings of the International Conference for High Performance Computing, Networking, Storage and Analysis},
articleno = {99},
numpages = {17},
location = {Atlanta, Georgia},
series = {SC '20},
url = {https://dl.acm.org/doi/10.5555/3433701.3433833}
}

@misc{nvidia_a100_datasheet_2020,
  author       = {{NVIDIA Corporation}},
  title        = {{NVIDIA A100} Tensor Core {GPU} Datasheet},
  year         = {2020},
  month        = jun,
  howpublished = {Datasheet},
  url          = {https://www.nvidia.com/content/dam/en-zz/Solutions/Data-Center/a100/pdf/nvidia-a100-datasheet.pdf},
  note         = {Accessed: 2026-04-08}
}

@misc{nvidia_h100_datasheet_2022,
  author       = {{NVIDIA Corporation}},
  title        = {{NVIDIA H100 Tensor Core GPU} Datasheet},
  year         = {2022},
  month        = sep,
  howpublished = {Datasheet},
  url          = {https://resources.nvidia.com/en-us-hopper-architecture/nvidia-tensor-core-gpu-datasheet},
  note         = {Accessed: 2026-04-08. Document No. 2287922}
}

@inproceedings{DBLP:conf/ics/SelvitopiBNTYB21,
  author       = {Oguz Selvitopi and
                  Benjamin Brock and
                  Israt Nisa and
                  Alok Tripathy and
                  Katherine A. Yelick and
                  Aydin Bulu{\c{c}}},
  editor       = {Huiyang Zhou and
                  Jose Moreira and
                  Frank Mueller and
                  Yoav Etsion},
  title        = {Distributed-memory parallel algorithms for sparse times tall-skinny-dense
                  matrix multiplication},
  booktitle    = {{ICS} '21: 2021 International Conference on Supercomputing, Virtual
                  Event, USA, June 14-17, 2021},
  pages        = {431--442},
  publisher    = {{ACM}},
  year         = {2021},
  url          = {https://doi.org/10.1145/3447818.3461472},
  doi          = {10.1145/3447818.3461472},
  bibsource    = {dblp computer science bibliography, https://dblp.org}
}

@inproceedings{DBLP:conf/sc/LagunaMMRSS19,
  author       = {Ignacio Laguna and
                  Ryan J. Marshall and
                  Kathryn M. Mohror and
                  Martin Ruefenacht and
                  Anthony Skjellum and
                  Nawrin Sultana},
  editor       = {Michela Taufer and
                  Pavan Balaji and
                  Antonio J. Pe{\~{n}}a},
  title        = {A large-scale study of {MPI} usage in open-source {HPC} applications},
  booktitle    = {Proceedings of the International Conference for High Performance Computing,
                  Networking, Storage and Analysis, {SC} 2019, Denver, Colorado, USA,
                  November 17-19, 2019},
  pages        = {31:1--31:14},
  publisher    = {{ACM}},
  year         = {2019},
  url          = {https://doi.org/10.1145/3295500.3356176},
  doi          = {10.1145/3295500.3356176},
  bibsource    = {dblp computer science bibliography, https://dblp.org}
}

@inproceedings{DBLP:conf/asplos/DingHZL0Y0P26,
  author       = {Yaoyao Ding and
                  Bohan Hou and
                  Xiao Zhang and
                  Allan Lin and
                  Tianqi Chen and
                  Cody Hao Yu and
                  Yida Wang and
                  Gennady Pekhimenko},
  editor       = {Benjamin C. Lee and
                  Harry Xu and
                  Mark Silberstein and
                  Bingyao Li},
  title        = {Tilus: {A} Tile-Level {GPGPU} Programming Language for Low-Precision
                  Computation},
  booktitle    = {Proceedings of the 31st {ACM} International Conference on Architectural
                  Support for Programming Languages and Operating Systems, Volume 1,
                  {ASPLOS} 2026, Pittsburgh, PA, USA, March 22-26, 2026},
  pages        = {281--297},
  publisher    = {{ACM}},
  year         = {2026},
  url          = {https://doi.org/10.1145/3760250.3762219},
  doi          = {10.1145/3760250.3762219},
  bibsource    = {dblp computer science bibliography, https://dblp.org}
}

@inproceedings{DBLP:conf/sc/JacquelinAM22,
  author       = {Mathias Jacquelin and
                  Mauricio Araya{-}Polo and
                  Jie Meng},
  editor       = {Felix Wolf and
                  Sameer Shende and
                  Candace Culhane and
                  Sadaf R. Alam and
                  Heike Jagode},
  title        = {Scalable Distributed High-Order Stencil Computations},
  booktitle    = {{SC22:} International Conference for High Performance Computing, Networking,
                  Storage and Analysis, Dallas, TX, USA, November 13-18, 2022},
  pages        = {30:1--30:13},
  publisher    = {{IEEE}},
  year         = {2022},
  url          = {https://doi.org/10.1109/SC41404.2022.00035},
  doi          = {10.1109/SC41404.2022.00035},
  bibsource    = {dblp computer science bibliography, https://dblp.org}
}

@inproceedings{DBLP:conf/cgo/LichtKMBHH21,
  author       = {Johannes de Fine Licht and
                  Andreas Kuster and
                  Tiziano De Matteis and
                  Tal Ben{-}Nun and
                  Dominic Hofer and
                  Torsten Hoefler},
  editor       = {Jae W. Lee and
                  Mary Lou Soffa and
                  Ayal Zaks},
  title        = {StencilFlow: Mapping Large Stencil Programs to Distributed Spatial
                  Computing Systems},
  booktitle    = {{IEEE/ACM} International Symposium on Code Generation and Optimization,
                  {CGO} 2021, Seoul, South Korea, February 27 - March 3, 2021},
  pages        = {315--326},
  publisher    = {{IEEE}},
  year         = {2021},
  url          = {https://doi.org/10.1109/CGO51591.2021.9370315},
  doi          = {10.1109/CGO51591.2021.9370315},
  bibsource    = {dblp computer science bibliography, https://dblp.org}
}

@article{DBLP:journals/taco/GysiMZHDWFHG21,
  author       = {Tobias Gysi and
                  Christoph M{\"{u}}ller and
                  Oleksandr Zinenko and
                  Stephan Herhut and
                  Eddie Davis and
                  Tobias Wicky and
                  Oliver Fuhrer and
                  Torsten Hoefler and
                  Tobias Grosser},
  title        = {Domain-Specific Multi-Level {IR} Rewriting for {GPU:} The Open Earth
                  Compiler for GPU-accelerated Climate Simulation},
  journal      = {{ACM} Trans. Archit. Code Optim.},
  volume       = {18},
  number       = {4},
  pages        = {51:1--51:23},
  year         = {2021},
  url          = {https://doi.org/10.1145/3469030},
  doi          = {10.1145/3469030},
  bibsource    = {dblp computer science bibliography, https://dblp.org}
}

@inproceedings{DBLP:conf/sc/RockiESSMKPDS020,
  author       = {Kamil Rocki and
                  Dirk Van Essendelft and
                  Ilya Sharapov and
                  Robert Schreiber and
                  Michael Morrison and
                  Vladimir Kibardin and
                  Andrey Portnoy and
                  Jean{-}Francois Dietiker and
                  Madhava Syamlal and
                  Michael James},
  editor       = {Christine Cuicchi and
                  Irene Qualters and
                  William T. Kramer},
  title        = {Fast stencil-code computation on a wafer-scale processor},
  booktitle    = {Proceedings of the International Conference for High Performance Computing,
                  Networking, Storage and Analysis, {SC} 2020, Virtual Event / Atlanta,
                  Georgia, USA, November 9-19, 2020},
  pages        = {58},
  publisher    = {{IEEE/ACM}},
  year         = {2020},
  url          = {https://doi.org/10.1109/SC41405.2020.00062},
  doi          = {10.1109/SC41405.2020.00062},
  bibsource    = {dblp computer science bibliography, https://dblp.org}
}

@inproceedings{MiyajimaSC25,
author = {Miyajima, Takaaki and Fukuoka, Leon},
title = {Benchmarking the Cerebras Wafer Scale Engine-2 Architecture},
year = {2025},
isbn = {9798400718717},
publisher = {Association for Computing Machinery},
address = {New York, NY, USA},
url = {https://doi.org/10.1145/3731599.3767438},
doi = {10.1145/3731599.3767438},
booktitle = {Proceedings of the SC '25 Workshops of the International Conference for High Performance Computing, Networking, Storage and Analysis},
pages = {818–822},
numpages = {5},
location = {
},
series = {SC Workshops '25}
}

@misc{ClaudeAI,
author = {
    Anthropic
},

year = {
    2025-2026
},

title = {
    Sonnet 4.5, {S}onnet 4.6
}

}

@misc{gpt54,
author = {
    OpenAI
},

year = {
    2026
},

title = {
    {GPT}-5.4
}

}

@inproceedings{DBLP:conf/popl/HondaYC08,
  author       = {Kohei Honda and
                  Nobuko Yoshida and
                  Marco Carbone},
  editor       = {George C. Necula and
                  Philip Wadler},
  title        = {Multiparty asynchronous session types},
  booktitle    = {Proceedings of the 35th {ACM} {SIGPLAN-SIGACT} Symposium on Principles
                  of Programming Languages, {POPL} 2008, San Francisco, California,
                  USA, January 7-12, 2008},
  pages        = {273--284},
  publisher    = {{ACM}},
  year         = {2008},
  url          = {https://doi.org/10.1145/1328438.1328472},
  doi          = {10.1145/1328438.1328472},
  bibsource    = {dblp computer science bibliography, https://dblp.org}
}

@inproceedings{DBLP:conf/pldi/Castro-Perez0GY21,
  author       = {David Castro{-}Perez and
                  Francisco Ferreira and
                  Lorenzo Gheri and
                  Nobuko Yoshida},
  editor       = {Stephen N. Freund and
                  Eran Yahav},
  title        = {Zooid: a {DSL} for certified multiparty computation: from mechanised
                  metatheory to certified multiparty processes},
  booktitle    = {{PLDI} '21: 42nd {ACM} {SIGPLAN} International Conference on Programming
                  Language Design and Implementation, Virtual Event, Canada, June 20-25,
                  2021},
  pages        = {237--251},
  publisher    = {{ACM}},
  year         = {2021},
  url          = {https://doi.org/10.1145/3453483.3454041},
  doi          = {10.1145/3453483.3454041},
  bibsource    = {dblp computer science bibliography, https://dblp.org}
}

@inproceedings{DBLP:conf/pdp/NgY14,
  author       = {Nicholas Ng and
                  Nobuko Yoshida},
  title        = {Pabble: Parameterised Scribble for Parallel Programming},
  booktitle    = {22nd Euromicro International Conference on Parallel, Distributed,
                  and Network-Based Processing, {PDP} 2014, Torino, Italy, February
                  12-14, 2014},
  pages        = {707--714},
  publisher    = {{IEEE} Computer Society},
  year         = {2014},
  url          = {https://doi.org/10.1109/PDP.2014.20},
  doi          = {10.1109/PDP.2014.20},
  bibsource    = {dblp computer science bibliography, https://dblp.org}
}

@inproceedings{DBLP:conf/cgo/Bronevetsky09,
  author       = {Greg Bronevetsky},
  title        = {Communication-Sensitive Static Dataflow for Parallel Message Passing
                  Applications},
  booktitle    = {Proceedings of the {CGO} 2009, The Seventh International Symposium
                  on Code Generation and Optimization, Seattle, Washington, USA, March
                  22-25, 2009},
  pages        = {1--12},
  publisher    = {{IEEE} Computer Society},
  year         = {2009},
  url          = {https://doi.org/10.1109/CGO.2009.32},
  doi          = {10.1109/CGO.2009.32},
  bibsource    = {dblp computer science bibliography, https://dblp.org}
}

@inproceedings{DBLP:conf/icalp/DenielouY13,
  author       = {Pierre{-}Malo Deni{\'{e}}lou and
                  Nobuko Yoshida},
  editor       = {Fedor V. Fomin and
                  Rusins Freivalds and
                  Marta Z. Kwiatkowska and
                  David Peleg},
  title        = {Multiparty Compatibility in Communicating Automata: Characterisation
                  and Synthesis of Global Session Types},
  booktitle    = {Automata, Languages, and Programming - 40th International Colloquium,
                  {ICALP} 2013, Riga, Latvia, July 8-12, 2013, Proceedings, Part {II}},
  series       = {Lecture Notes in Computer Science},
  volume       = {7966},
  pages        = {174--186},
  publisher    = {Springer},
  year         = {2013},
  url          = {https://doi.org/10.1007/978-3-642-39212-2\_18},
  doi          = {10.1007/978-3-642-39212-2\_18},
  bibsource    = {dblp computer science bibliography, https://dblp.org}
}

@misc{zig,
  author = {{Zig Software Foundation}},
  title = {The {Zig} Programming Language},
  howpublished = {https://ziglang.org/},
  year = {2026}
}

\end{document}